\documentclass[sigconf, pageno,lineno, authorversion]{acmart}

\usepackage{xfrac}
\usepackage{flushend}
\usepackage{amsmath}
\usepackage{mathtools}
\AtBeginDocument{%
  \providecommand\BibTeX{{%
    \normalfont B\kern-0.5em{\scshape i\kern-0.25em b}\kern-0.8em\TeX}}}

\usepackage{algorithm}
\usepackage[noend]{algpseudocode}

\copyrightyear{2023} 
\acmYear{2023} 
\setcopyright{acmlicensed}\acmConference[e-Energy '23]{The 14th ACM International Conference on Future Energy Systems}{June 20--23, 2023}{Orlando, FL, USA}
\acmBooktitle{The 14th ACM International Conference on Future Energy Systems (e-Energy '23), June 20--23, 2023, Orlando, FL, USA}
\acmPrice{15.00}
\acmDOI{10.1145/3575813.3576870}
\acmISBN{979-8-4007-0032-3/23/06}

\makeatletter
\newcommand{\algmargin}{\the\ALG@thistlm}
\makeatother
\newlength{\whilewidth}
\algdef{SE}[parWHILE]{parWhile}{EndparWhile}[1]
{\parbox[t]{\dimexpr\linewidth-\algmargin}{%
		\hangindent\whilewidth\strut\algorithmicwhile\ #1\ \algorithmicdo\strut}}{\algorithmicend\ \algorithmicwhile}%
\algdef{SE}[SUBALG]{Indent}{EndIndent}{}{\algorithmicend\ }%
\algtext*{Indent}
\algtext*{EndIndent}
\algnewcommand{\parState}[1]{\State%
	\parbox[t]{\dimexpr\linewidth-\algmargin}{\strut #1\strut}}
\makeatother

\newtheorem{theorem}{Theorem}

\newtheorem*{lemma*}{Lemma}

\newtheorem{definition}[theorem]{Definition}

\begin{document}

\title[Equitable Network-Aware Decarbonization of Residential Heating at City Scale]{Equitable Network-Aware Decarbonization of Residential Heating at City Scale}



\author{Adam Lechowicz}
\affiliation{%
  \institution{University of Massachusetts Amherst}
    \country{}
}
\email{alechowicz@cs.umass.edu}

\author{Noman Bashir}
\affiliation{%
  \institution{University of Massachusetts Amherst}
    \country{}
}
\email{nbashir@cs.umass.edu}

\author{John Wamburu}
\affiliation{%
  \institution{IBM Research Africa}
    \country{}
}
\email{john.wamburu@ibm.com}

\author{Mohammad Hajiesmaili}
\affiliation{%
  \institution{University of Massachusetts Amherst}
    \country{}
}
\email{hajiesmaili@cs.umass.edu}

\author{Prashant Shenoy}
\affiliation{%
  \institution{University of Massachusetts Amherst}
    \country{}
}
\email{shenoy@cs.umass.edu}

\renewcommand{\shortauthors}{Lechowicz, Bashir, Wamburu, Hajiesmaili, Shenoy}

\begin{abstract}


Residential heating, primarily powered by natural gas, accounts for a significant portion of residential sector energy use and carbon emissions in many parts of the world.
Hence, there is a push towards decarbonizing residential heating by transitioning to energy-efficient heat pumps powered by an increasingly greener and less carbon-intensive electric grid. 
However, such a transition will add additional load to the electric grid triggering infrastructure upgrades, and subsequently erode the customer base using the gas distribution network.
Utilities want to guide these transition efforts to ensure a phased decommissioning of the gas network and deferred electric grid infrastructure upgrades while achieving carbon reduction goals. 
To facilitate such a transition, we present a network-aware optimization framework for decarbonizing residential heating at city scale with an objective to maximize carbon reduction under budgetary constraints. 
Our approach operates on a graph representation of the gas network topology to compute the cost of transitioning and select neighborhoods for transition. 
We further extend our approach to explicitly incorporate equity and ensure an equitable distribution of benefits across different socioeconomic groups. 
We apply our framework to a city in the New England region of the U.S., using real-world gas usage, electric usage, and grid infrastructure data.
We show that our \emph{network-aware} strategy achieves 55\% higher carbon reductions than prior network-oblivious work under the same budget.
Our \emph{equity-aware} strategy achieves an equitable outcome while preserving the carbon reduction benefits of the \emph{network-aware} strategy.

\end{abstract}





\maketitle

\section{Introduction}
\label{sec:intro}


Residential heating is responsible for roughly 50\% of total residential energy use in colder climates.  Large parts of the world depend on fossil fuels such as oil, propane, or natural gas for heating homes when outdoor temperatures are low.  In the U.S., residential energy use contributes about 20\% of the total yearly greenhouse gas emissions \cite{Goldstein:2020}.  The decarbonization of residential heating systems is therefore an important goal for a carbon-free future.


A promising approach to address this problem is to electrify heating systems by replacing fossil fuel-based heating components such as furnaces and boilers with electric heat pumps.  
Heat pump technology has advanced significantly over the past decade, enabling drop-in, air-source units which maintain good efficiency even at temperatures below -15$^{\circ}$C~\cite{Johnson:2021, Schoenbauer:2016}, making them suitable for cold climates.  
From an efficiency perspective, heat pumps use less energy compared to fossil fuel combustion to output the same amount of heat~\cite{OEERE}.  
Combined with the ability to use electricity from carbon-free and renewable sources, heat pumps offer an attractive solution for energy efficiency and reduced carbon emissions.

While there are substantial decarbonization benefits to electrifying residential heating, such a transition has significant implications for the gas and electric utilities.  
As residential customers progressively switch from fossil-based heating to electric heat pumps, the customer base of a gas utility will shrink over time.  
Since a gas distribution network which serves a dwindling customer base is not cost-effective, a \emph{phased decommissioning} of its gas network is desired -- switching entire neighborhoods to electric heating and shutting down the gas distribution in those neighborhoods.
Under this scenario, an electric utility will face the opposite problem -- as more homes switch to electric heating, demand on the distribution grid will increase. 
The utility may need to upgrade the electric grid infrastructure (e.g. transformers) to serve the additional demand.

Prior work has studied decarbonization of heating and electrification from a homeowner's perspective, where homeowners are incentivized, based on their carbon emissions or carbon inefficiency, to convert to energy-efficient electric heat pumps \cite{john-buildsys22}.  However, such approaches encourage homeowners to make independent decisions which are agnostic of the gas network's topology or electric grid considerations.  
While they reduce overall emissions, they may not enable gas utilities to reduce maintenance costs, especially if the electrified homes are scattered throughout a city and prevent the utility from decommissioning its distribution network.

A utility-based approach where a transition to heat pumps is realized in a planned manner, converting entire neighborhoods at a time, can yield significant benefits for both the gas and electric utilities.
The gas utility can shut down entire portions of its distribution network, reducing maintenance costs, while the electric utility can handle additional demand using targeted infrastructure upgrades.  
Such an approach requires a network-aware methodology that leverages network topology information to makes strategic decisions while selecting households or neighborhoods for transition.
Furthermore, such interventions should ensure that the benefits of transition are equitably distributed across socioeconomic groups and do not perpetuate the existing inequalities of our society. 

To facilitate such a transition, we develop an optimization framework that maximizes carbon reductions in an equitable manner for a given budget. 
In doing so, we make the following contributions.

\smallskip

\noindent \textbf{Network-aware optimization framework.} 
We present a network-aware optimization framework that maximizes carbon emission reductions while transitioning a subset of neighborhoods to electric heat pumps under budgetary constraints. We devise a tailored solution for solving the optimization problem by addressing several design challenges that arise from leveraging network topology information to ensure a cost-effective transition. 

\smallskip 

\noindent \textbf{Equity-aware optimization.}
We extend our \emph{network-aware} \mbox{optimization} framework, incorporating equitable constraints which set specific budget allocations for different socioeconomic groups to ensure that transition investments do not exacerbate inequities. 
In our case study, the \emph{equity-aware} strategy transitions 14.3\% more homes than \emph{network-aware} strategy, primarily from low- and middle-income tracts to achieve an equitable budget distribution.

\noindent \textbf{Evaluation using real-world case study.}
We evaluate our \emph{network-aware} and \emph{equity-aware} strategies using real-world usage data (gas \& electric) and infrastructure information from a city in New England, USA.  Our \emph{network-aware} strategy achieves 55\% higher carbon reduction compared to prior network-oblivious work, and the \emph{equity-aware} strategy preserves these carbon reduction benefits while ensuring an equitable distribution of the budget.

\section{Background}
\label{sec:background}

In this section, we present background on natural gas distribution, the electric grid, and air-source electric heat pumps.

\smallskip

\noindent\textbf{Natural gas heating and distribution.}
Natural gas is a primary fuel used for residential heating in many parts of the world.  In the U.S. alone, gas represents over 52\% of heating energy use \cite{EIA:2015}. In heating applications, natural gas fuels a burner which heats air or water in furnace or boiler systems, respectively. The natural gas combustion releases greenhouse gases such as CO$_2$ into the atmosphere as a byproduct. In addition, on average, 2.3\% of the natural gas produced in the U.S. leaks into the atmosphere \cite{Alvarez:2018} -- natural gas contains methane, which is a greenhouse gas at least 25 times more potent than CO$_2$ \cite{EPA:Methane}. A transition away from the usage of natural gas for home heating is a critical step towards achieving greenhouse gas (GHG) emissions goals. 

The primary mode of delivery for natural gas is an expansive pipeline system which connects gas wells, refineries, compressor stations, and eventually gas consumers.  This pipeline network comprises transmission pipelines, which move large amounts of high pressure gas over long distances, and distribution lines, which are managed by a utility and connect the end users to a low-pressure supply.  Also known as ``mains’’, distribution lines run through neighborhoods and supply gas through ``service'' lines, which connect to a gas meter at the point of use.  The meter measures a building’s gas usage by end uses, such as stoves or furnaces.

In this paper, we focus on the network of distribution lines at city scale, which may have hundreds of kilometers in pipes, and consume a sizable portion of a utility's maintenance budget. 
In the context of the energy transition, the utility is incentivized to reduce this maintenance expense in the near term in order to redirect resources elsewhere -- this corresponds with decommissioning natural gas infrastructure wherever possible.
The layout of the gas network and the order in which they are chosen for transition are important considerations while pursuing long-term phase-out of the gas network. 
For example, it can be beneficial for the utility to retrofit homes in neighborhoods where entire distribution lines can be decommissioned. An alternative strategy can pick homes where the additional electricity demand of heat pumps would not trigger an electric grid upgrade, a key consideration for utilities that own both the gas and electric networks. 
Our work explores answering similar questions under budgetary constraints.


\smallskip

\noindent\textbf{Electric distribution grid.}
\label{sec:distgrid}
This work focuses on the distribution part of the electric grid, which supplies electricity to end consumers. 
The distribution grid encompasses substations, feeders, and transformers, which can be viewed as a hierarchical network~\cite{EIA:2016}.  
We base our analysis on transformers that exist at the edge of the distribution network and connect directly to the end users.  
Since we are interested in heating loads specifically, we ignore the purpose-built portions of the network which serve industrial customers such as manufacturing plants.
We assume that edge transformers serving residential spaces or small commercial spaces are likely to shoulder the burden of increasing electric heating loads.

Edge transformers significantly vary in their capacity, ranging from small pole-top 5-10 kilo-Volt-Ampere (kVA)
to larger 75, 100, and 225 kVA transformers. 
Smaller pole-top units may serve a few homes (e.g. 2 to 4), while larger ones may serve apartment complexes, commercial spaces, or office buildings.
Utilities size edge transformers based on the expected peak load for a particular location. 
Transformers can operate above their rated capacities if the actual load exceeds the expected peak, but they will generate excessive heat which is absorbed by safety mechanisms such as mineral oil. 
While transformers can operate at up to 125\% of their rated capacity~\cite{Jardini:97}, this scenario is not desirable as it decreases the transformer efficiency and can cause the insulating oil to evaporate, leading to melting of coils and power outages.

\smallskip

\noindent\textbf{Air-source electric heat pumps.}
Air-source electric heat pumps (ASHPs) offer an energy-efficient alternative to gas-based heating.  
ASHPs leverage the latent heat of vaporization to concentrate thermal energy and physically transfer it, which is used to heat an indoor space.  
By \emph{moving} heat instead of \emph{generating} it, heat pumps can be more energy-efficient than heating technologies which rely on generation, such as furnaces, boilers, and resistive heating.  
Moreover, as electricity is increasingly sourced from carbon-free sources, ASHPs can significantly reduce the carbon footprint of heating in the long run.
The efficiency of ASHPs changes as a function of outdoor temperature because they rely on air heat transfer. As the outside temperature decreases, ASHP heating capacity decreases. In the past, heat pumps required supplementary equipment such as a gas furnace or resistive heating for deployments in cold climates~\cite{Gschwend:2016}. However, recent advances in heat pump technology make them a viable solution for low temperature regions~\cite{Schoenbauer:2016, Dymond:2018}. 

\section{Problem and Methodology}
\label{sec:prob}

\begin{table}[t]
	\caption{A summary of key notations }
	\vspace{-4mm}
	\label{tab:notations}
 	\footnotesize
	\begin{center}
		\begin{tabular}[P]{|c|l|c|}
			\hline
			\textbf{Notation} & \textbf{Description} & \textbf{Icon}\\
			\hline \hline
            $G = (\mathcal{V}, \mathcal{E})$ & Directed graph representing gas network &\\
			\hline
            $(u,v) \in \mathcal{E}$ & Edge from node $u$ to node $v$ in $G$ &\\
			\hline
            $m(u,v)$ & Maintenance cost of edge $(u,v)$ &\begin{minipage}{5mm}
                \includegraphics[width=5mm, height=5mm]{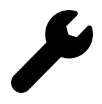}
            \end{minipage}\\
			\hline
            $s \in \mathcal{V}$ & Source node for gas distribution &\\
            \hline
            $\mathcal{N}$ & Set of neighborhoods $\mathcal{N}$ &\begin{minipage}{6mm}
                \includegraphics[width=6mm, height=4.5mm]{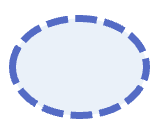}
            \end{minipage}
            \\
			\hline\hline
			$\mathcal{H}$ & The set of households, indexed by $j$ &\begin{minipage}{5mm}
                \includegraphics[width=5mm, height=5mm]{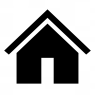}
            \end{minipage}\\
			\hline
            $g(j)$  & Natural gas usage for $j$th house &\begin{minipage}{5mm}
                \includegraphics[width=5mm, height=5mm]{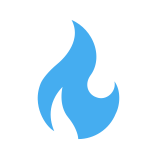}
            \end{minipage}\\
			\hline
            $p(j)$ & ASHP installation cost for $j$th house &\\
			\hline\hline
			$\mathcal{T}$ & The set of transformers, indexed by $i$  &\begin{minipage}{5mm}
                \includegraphics[width=4.5mm, height=4.5mm]{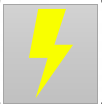}
            \end{minipage}\\
			\hline
            $t(j)$ & Transformer supplying $j$th house &\begin{minipage}{5mm}
                \includegraphics[width=5mm, height=3mm]{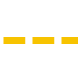}
            \end{minipage}\\
			\hline
            $c(i)$ & Capacity rating (kVA) for $i$th transformer &\\
			\hline
            $L^{\texttt{peak}}(i)$ & Peak load for $i$th transformer &\\
			\hline
            $r(i)$ & Replacement cost for $i$th transformer &\\
			\hline
		\end{tabular}
	\end{center}
	\vspace{-0.70cm}
\end{table}

In this section, we develop a general optimization framework from a utility's perspective with a limited budget for decarbonization. Given the budget constraints, the objective is to maximize the carbon reduction of residential heating while transitioning from gas-based heating to air-source electric heat pumps.  
Our framework takes a network-aware approach for a city-scale decarbonization and captures the impact of transition on the electricity distribution network. 
We also extend our optimization framework to incorporate equitable allocation in the transition process to avoid perpetuating existing socioeconomic inequities.




\vspace{-0.15cm}
\subsection{System Model}
\label{sec:sys_mod}

\begin{figure}[t]
    \centering
    \includegraphics[width=0.8\columnwidth]{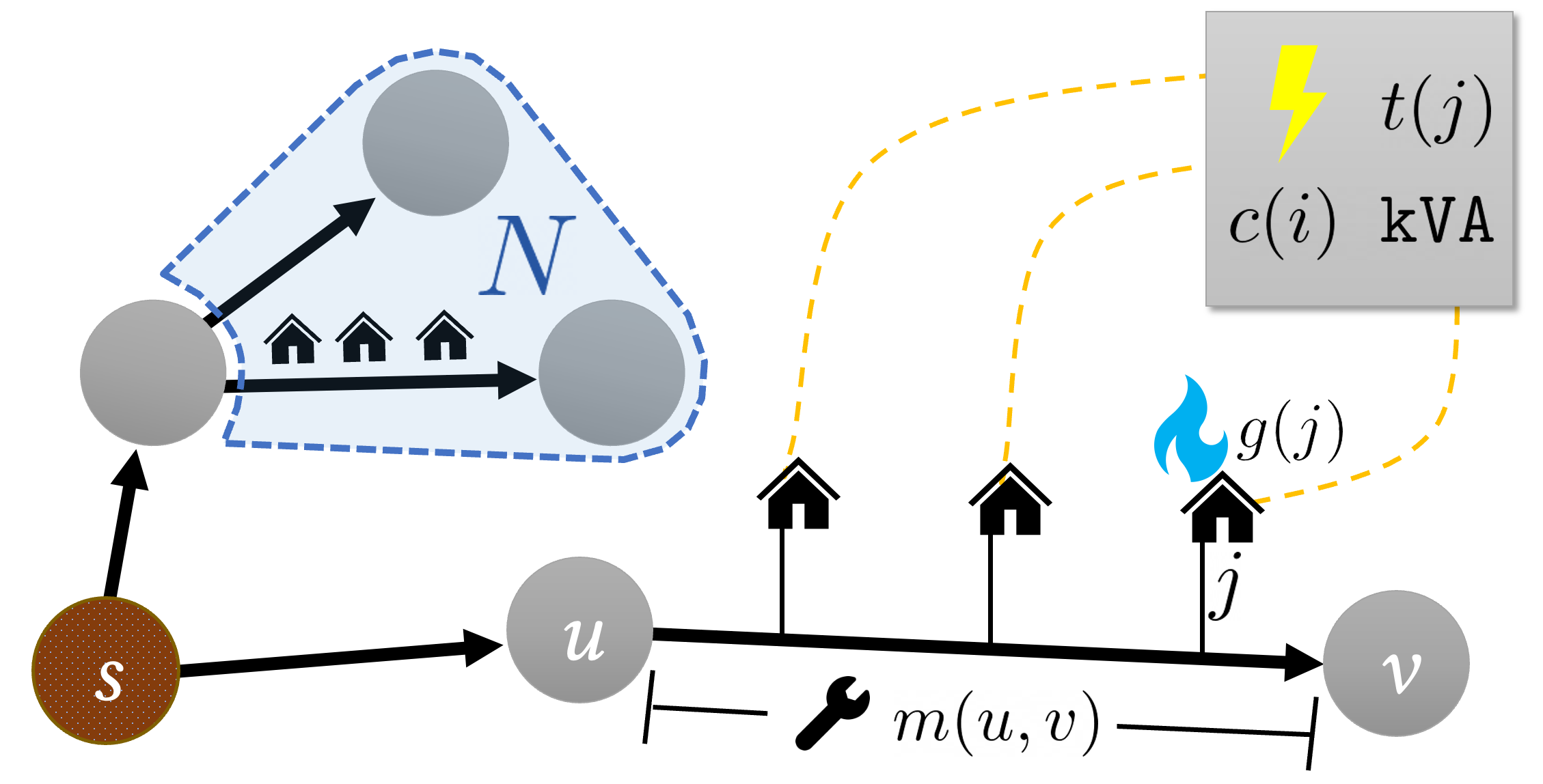}
    \vspace{-0.3cm}
    \caption{A graphical representation of system model.}
    \label{fig:systemIllustration}
    \vspace{-0.6cm}
\end{figure}

In this section, we introduce the system model. 
We outline key notations in Table~\ref{tab:notations} and illustrate the model with its components in Figure~\ref{fig:systemIllustration}.


Let $G = (\mathcal{V}, \mathcal{E})$ be a directed graph representing a city-scale gas distribution network.  
This graph captures the directed flow characteristic of a real gas distribution network.
We consider $G$ to be a \emph{weakly connected} graph, such that there exist paths between some source node $s \in \mathcal{V}$ 
and all other nodes in the network. 

We have a set of households $\mathcal{H}$ that are served by the gas network $G$ and the electric grid through a set of electric transformers $\mathcal{T}$.
Each transformer $i \in \mathcal{T}$ has a capacity $c(i)$ and a peak load $L^{\texttt{peak}}(i)$.  Note that the peak load of the transformer $i$ only changes when houses served by that transformer are transitioned to ASHPs.

We define several attributes for each house $j \in \mathcal{H}$.
Each household has a gas usage profile denoted by $g(j)$.  In order to convert this gas usage into an equivalent ASHP electric load, we can convert $g(j)$ into the corresponding measure of heat energy.  Given this quantity of heat energy, finding the electric load required to generate a given amount of heat depends on the ASHP's COP (coefficient of performance) and a unit conversion.  
This additive electric load is supplied by the transformer serving house $j$, denoted by $t(j)$. The estimated cost, after subsidies and rebates, for the utility to install an ASHP system in house $j$ is represented by $p(j)$.  Finally, we define a constant $E^{\texttt{C0$_2$}}$ which scales a given gas usage $g(j)$ to obtain the CO$_2$ emissions released as a result of combustion.

Next, each edge $(u,v) \in \mathcal{E}$, represents a gas pipeline and stores several attributes such as its material, length, age, and service lines connected to it.
Let $\mathcal{H}_{u,v}$ denote the homes connected to the gas pipeline $(u,v)$.
The representation of buildings as a part of the edge allows a granular control over the grid topology. 
For example, if the pipeline $(u,v)$ is shut down, each house assigned to that edge must be transitioned away from natural gas.  
Such a scenario is cost-effective and desirable for the utility as it allows shutting down an entire gas pipeline once all of the connected homes have been transitioned~\cite{PaloAlto:2020}.  
The potential savings from the suspended future maintenance cost of pipeline segment $(u,v)$ are given by $m(u,v)$.

We define a set of \emph{neighborhoods} $\mathcal{N}$, such that each $N \in \mathcal{N}$ is a subset of $\mathcal{E}$.  
The exact composition of $\mathcal{N}$ is arbitrary and can be defined separately -- under a restriction that each neighborhood must be a set of edges from $\mathcal{E}$.  
In this problem, the optimization seeks to find a subset of neighborhoods $\mathcal{C} \subseteq \mathcal{N}$ which, when transitioned to ASHPs, maximize a given objective function subject to the budget constraint.  
After transition, $\mathcal{C}$ constitutes a set of edges $\mathcal{E}_{\mathcal{C}}$ where gas service has been shut down and a set of converted households $\mathcal{H}_{\mathcal{C}}$. 
We also define a set $\mathcal{T}_{\mathcal{C}} \subseteq \mathcal{T}$ which denotes \emph{newly overloaded transformers} post transition.   
Each overloaded transformer $i \in \mathcal{T}_{\mathcal{C}}$ includes an upgrade and replacement cost $r(i)$, which captures the cost of electric grid upgrades due to additional electricity demand.

\smallskip
\noindent{\bf Budgetary constraints.}
The total budget $B$ places a constraint on the network-aware heat pump transition expenses.
The total cost of transition has three key components. 
First is the total cost of installing ASHPs in houses selected for transition, computed for the converted homes as follows:
\begin{align}
    \sum_{j \in \mathcal{H}_{\mathcal{C}}} p(j).
\end{align}
Second component is the cost of electric grid upgrades due to additional demand from heat pumps, computed for the newly overloaded transformers as follows:
\begin{align}
    \sum_{i \in \mathcal{T}_{\mathcal{C}}} r(i).
\end{align}
The final component is the \emph{savings} on maintenance costs attributed to gas pipeline shutdowns, which is given by a summation over the edges inside the converted neighborhoods:

\vspace{-0.4cm}
\begin{align}
    \sum_{(u,v) \in \mathcal{E}_{\mathcal{C}}} m(u,v).
\end{align}

\vspace{-0.15cm}
The total cost of the network-aware heat pump conversion is then given by the following expression:

\vspace{-0.4cm}
\begin{align}
    \;\sum_{\mathclap{j \in \mathcal{H}_{\mathcal{C}}}}\; p(j) \; + \; \;\sum_{\mathclap{i \in \mathcal{T}_{\mathcal{C}}}}\; r(i) \; - \;\sum_{\mathclap{(u,v) \in \mathcal{E}_{\mathcal{C}}}}\; m(u,v),
\end{align}

\subsection{Optimization Formulation}
\label{sec:opt}

We now summarize the full formulation of the Network-aware Heat Pump Transition (\texttt{NHPT}) problem as an optimization.
\begin{align}
    \textnormal{max} \; & \;\sum_{\mathclap{j \in \mathcal{H}_{\mathcal{C}}}}\; E^{\texttt{C0$_2$}} g(j), \;\;\;\;\;\; \triangleright \textit{carbon emissions reduction} \label{eq:obj} \\
    \textnormal{s. t.,} \; & \;\sum_{\mathclap{j \in \mathcal{H}_{\mathcal{C}}}}\; p(j) \; + \;\sum_{\mathclap{i \in \mathcal{T}_{\mathcal{C}}}}\; r(i) \; - \;\sum_{\mathclap{(u,v) \in \mathcal{E}_{\mathcal{C}}}}\; m(u,v) \; \leq \; B. \label{eq:cons}
\end{align}
where the objective is to maximize the reduction in network-wide carbon emissions subject to budgetary constraints enforced by Eq.~\eqref{eq:cons}.   
Note that this formulation does not consider the potential of carbon emissions due to the increased electric load.  The average annual carbon intensity of the grid is assumed to be the same throughout the city. While this assumption does not capture the reality of different carbon intensities due to PV installations, power dispatch, and other factors, the optimization considers the annual average carbon emissions for each house, making this a useful simplifying assumption -- since we assume that the grid impacts each household's carbon uniformly, carbon due to electric load does not impact the optimization.  For a precise application, this additive carbon from electricity could be factored into the carbon constant $E^{\texttt{C0$_2$}}$, slightly reducing the net carbon reduction.


\subsection{Equitable Allocation}\label{sec:equityDef}
Our framework, described in 
Eq. \eqref{eq:obj} \& \eqref{eq:cons}, does not impose any restrictions on how the overall budget is distributed across the population. 
In optimizing for carbon emission reduction, it could disproportionately select affluent households which tend to have higher gas usage and higher emissions~\cite{WamburuGrazierIrwinCragoShenoy:2022}. 
As a result, an equity-agnostic transition can exacerbate existing societal inequities by increasing the energy efficiency of wealthy households and ignoring economically-disadvantaged households. 
In addition, a transition strategy which is perceived as inequitable can face greater challenges being accepted by constituent communities, ultimately impacting the achievable carbon reduction.
To facilitate \emph{equity-aware} transition, we extend our framework to ensure that different socio-economic groups receive an equitable share of the budget. 
These groups can be based on various factors including income, house value, and demographics. 
Equitable allocation should also maintain a high level of performance with regards to carbon reduction and environmental impact.


Suppose there are $K$ constituent groups, and a \emph{desired allocation} vector $\vec{D}$.  Each $k$th entry of $\vec{D}$ represents the desired allocation for group $g_k$, with respect to the total budget $B$:
\begin{align}
    \vec{D} &= \langle d_1, d_2, d_3, ... , d_{K} \rangle , \;\;\;\;\;\;\;\; \sum_{k=1}^K d_k = B.
\end{align}
Given a selected subset of neighborhoods $\mathcal{C}$, we define each $\mathcal{C}_k \subseteq \mathcal{C}$ to represent the subset of selected neighborhoods which belong to the $k$th group.

We formalize the Equitable Network-aware Heat Pump Transition (\texttt{E-NHPT}) problem as an extension to the \texttt{NHPT} optimization:
\vspace{-0.4cm}
\begin{align}
    \textnormal{max} \; &  \;\sum_{\mathclap{j \in \mathcal{H}_{\mathcal{C}}}}\; E^{\texttt{C0$_2$}} g(j), \;\;\;\;\;\; \triangleright \textit{carbon emissions reduction} \\
    & \sum_{\mathclap{j \in \mathcal{H}_{\mathcal{C}_k}}} p(j) + \sum_{\mathclap{i \in \mathcal{T}_{\mathcal{C}_k}}} r(i) - \;\sum_{\mathclap{(u,v) \in \mathcal{E}_{\mathcal{C}_k}}}\; m(u,v) \leq d_k, k \in \{1,2,..., K\}.
\end{align}
The objective of \texttt{E-NHPT} remains to maximize the reduction of network-wide carbon emissions -- the additional constraints serve as a \emph{guarantee} which define the amount of allocation a given group can obtain in any solution.  While the LP formulation includes each $d_k$ as a ``cap'' on the allocated budget to each group, the desired allocation vector $\vec{D}$ also implies an exact allocation for each group -- at the end of the process, if we have used nearly all of the total budget, then we know that each group has been allocated almost exactly $d_k$.
The \emph{equity-aware} optimization may not yield the absolute optimal savings, but allows for transition process that is equitable across socio-economic groups even at a city-scale.

\section{Solution Design}
\label{sec:design}
 
\begin{figure}[t]
    \vspace{-1em}
    \hfill
    \minipage{0.8\columnwidth}
        \includegraphics[width=\linewidth]{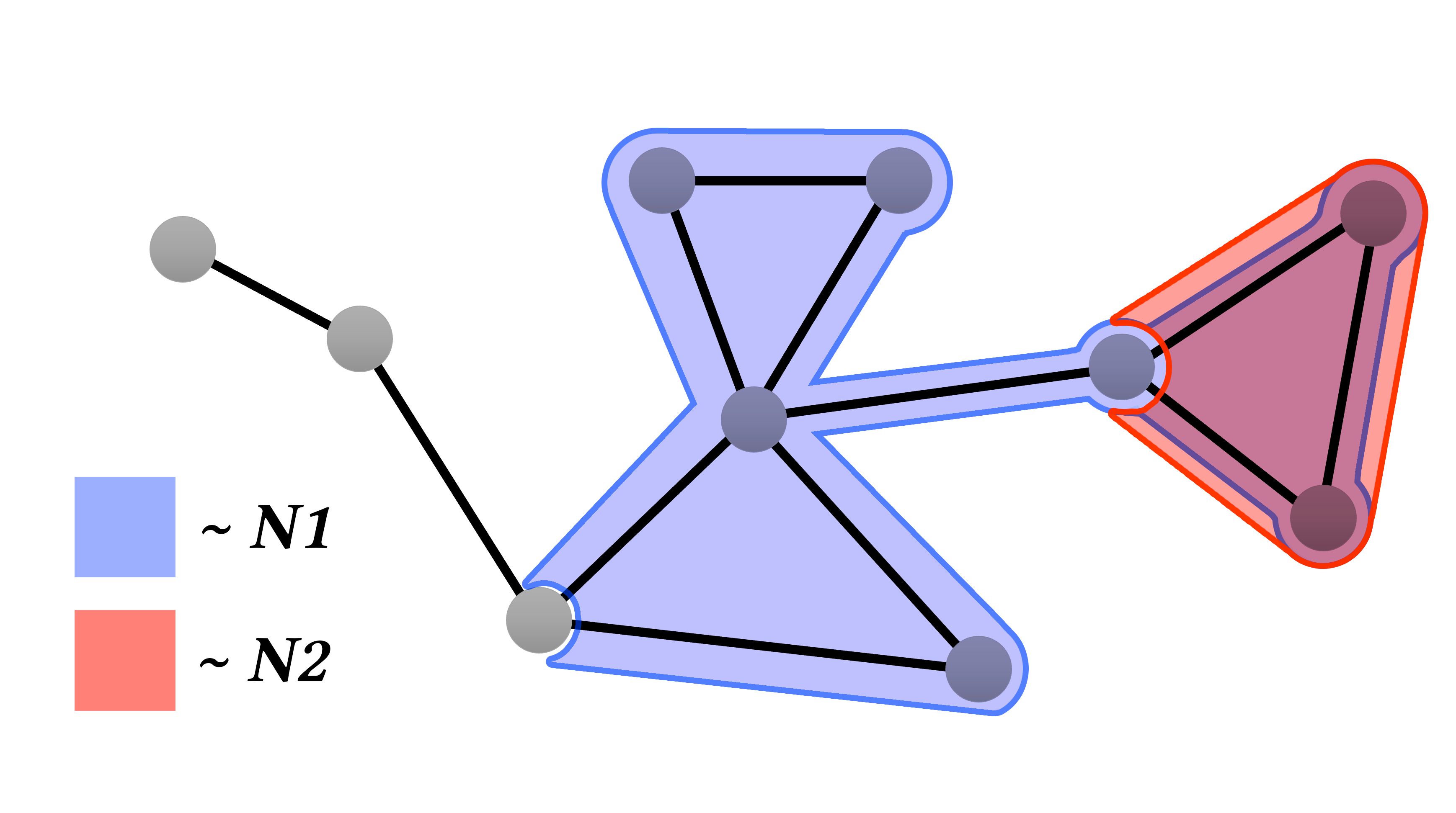}
    \endminipage
    \hfill
    \vspace{-0.75cm}
    \caption{An example graph with intersecting ``neighborhoods'', $N1$ and $N2$, where $N2$ (red) is a subset of $N1$ (blue).}
    \label{fig:graphExample}
    \vspace{-0.5cm}
\end{figure}

In this section, we first highlight the algorithmic challenges for solving \texttt{NHPT}.
Then, we present a procedure to solve the \texttt{NHPT} problem by considering the set of candidate neighborhoods as a knapsack problem. 

Finally, we present a methodological extension to the knapsack problem which explicitly considers the allocation to each group -- this extension can be used as a subroutine to solve the \texttt{E-NHPT} \textit{equitable allocation} problem.

\subsection{Algorithmic Challenges}
While the formulation of the \texttt{NHPT} problem is simple, algorithmic challenges associated with the topology of the gas network necessitate a tailored solution design.  
Consider the simple example illustrated in Figure~\ref{fig:graphExample}, which shows two highlighted neighborhoods in an arbitrary example graph.  
Note that neighborhood $N2$ is a subset of neighborhood $N1$.  
This can cause complexities in a sequential setting where either $N1$ or $N2$ is ``converted'' before the other -- for instance, if $N1$ is converted, $N2$ is trivially converted as well.  
Concretely, a crucial challenge is that neighborhoods are not necessarily independent of each other -- each neighborhood's parameters may depend on the status of an arbitrary number of other neighborhoods.  
This reflects how the graph's structure imposes additional ``hidden dependency constraints'' on the optimization problem, which are difficult to formalize as part of the problem. Hence, it is intractable to use an optimization solver to find a solution for \texttt{NHPT}. Instead, in \S\ref{sec:design}, we present a problem-specific solution design, which directly captures the dependency constraints in decision making. 


\subsection{Estimating Cost of Transition}
\label{sec:costNeigh}

To compare different neighborhoods and evaluate their suitability for transition, we compute the estimated cost of transition and the estimated carbon emission reductions from the transition. 
These values and their ratio, termed as the utility of a neighborhood, are used by our subsequent optimization solution.



\smallskip
\noindent\textbf{Neighborhood transition cost.}
For each neighborhood, we compute the total cost of transition to heat pumps, which consist of ASHP installation cost, grid upgrade cost, and savings obtained from gas network shutdowns.  
We capture the transition cost over all the households, edges, and transformers in $N$ as follows:
\begin{align}
    \text{TotalCost}_N &= \sum_{j \in N} p(j) \; + \sum_{i \in \mathcal{T}_{N}} r(i) \; - \sum_{(u,v) \in \mathcal{E}_{N}} m(u,v).
\end{align}

\vspace{-0.15cm}
\noindent\textbf{Neighborhood carbon reduction.}
We have the constant $E^{\texttt{C0$_2$}}$ which scales a given gas usage $g(j)$ to obtain the CO$_2$ emissions released as a result of combustion.
The exact value of $E^{\texttt{C0$_2$}}$ depends on the units used for the gas usage and carbon emissions.
Given this, the following sum over carbon emissions for each household in $N$ yields the carbon reduction of $N$:
\begin{align}
    \text{CarbonReduction}_N &= \sum_{j \in N} E^{\texttt{C0$_2$}} g(j).
\end{align}



\noindent\textbf{Neighborhood utility.}
For a neighborhood $N$, we consider the carbon reduction as the transition \emph{value} and the total cost as the transition \emph{weight}. 
We term the \emph{value-to-weight} ratio as the \emph{utility} of neighborhood $N$, defined as $\frac{\text{CarbonReduction}_N}{\text{TotalCost}_N}$.  Neighborhoods with higher \emph{utility} achieve more carbon reduction for a given cost. 
    

\smallskip

\noindent\textbf{Additional considerations.}
\label{sec:recomputation}
The cost and emission estimates are straightforward and easy to compute.
However, the topology of the gas network introduces complexities that complicate the seemingly trivial task. 
Let's consider the example from Fig. \ref{fig:graphExample} -- utility estimates can break down if either $N1$ or $N2$ is transitioned before the other.  
For example, if $N2$ is transitioned before $N1$, the utility estimate for $N1$ becomes invalid as it double-counts for the households in $N2$ (note that $N2$ is a subset of $N1$).
Similarly, if $N1$ is transitioned before $N2$,  the utility estimate for $N2$ becomes invalid because $N2$ homes cannot be transitioned twice.

To counter this problem, we can recompute the utility of all the neighborhoods as the graph evolves post-transition. 
This increases the runtime of the solution, but the process could be accelerated by recomputing the utility at each iteration only for neighborhoods which are likely to be affected because of proximity to the previous gas shutdown.
Furthermore, in a fairly dense network with many neighborhood candidates, it is less likely that we will face these challenges during the selection process.

\subsection{Neighborhood Selection}

We frame the problem of selecting candidates as a knapsack problem, where each candidate neighborhood is an ``item'' with a value (carbon reduction) and a weight (estimated transition cost).
The objective is to maximize the value of items packed into the knapsack $(\mathcal{S})$ without violating the knapsack's capacity $B$.


Solving knapsack is a known NP-hard problem \cite{Karp:1972}, but has a pseudo-polynomial time dynamic programming solution.  
We use the dynamic programming approach to select neighborhoods for transition from a list of candidates, as described in Algorithm \ref{alg:knap}.  
There is prior work that proposes approximation algorithms with better time complexity for the knapsack problem \cite{Lawler:1977}. 
We plan to explore using such approximation algorithms 
in future work. 
While the knapsack solver provides a solution, it may not be optimal given the challenges discussed in Section \ref{sec:costNeigh}.
If any of the selected neighborhoods are \emph{not independent}, i.e. suppose that $\exists N1, N2 \in \mathcal{S} : (N1 \subset N2) \vee (N2 \subset N1)$, a better solution may exist as the ``value'' of one neighborhood could include the carbon reduction and cost of another neighborhood.


To address this, we select a fresh set of neighborhoods in each iteration for the remaining budget whenever an attempt to convert a neighborhood does not yield the expected utility.  
This can happen if the utility of the neighborhood under consideration has changed after transitioning other neighborhoods.
We also use a heuristic based on \emph{utility} to sequentially convert the selected neighborhoods in the order of carbon reduction with respect to cost.  
Proving the optimality of this approach is non-trivial and we plan to further investigate algorithmic solutions to address this challenge.

\newlength{\textfloatsepsave} 
\setlength{\textfloatsepsave}{\textfloatsep} 
\setlength{\textfloatsep}{-10pt} 
\begin{algorithm}[t]
	\caption{Dynamic Programming Knapsack}
	\label{alg:knap}
	\begin{algorithmic}[1]
		\State \textbf{input:} capacity $B$, list of item weights $w$, list of item values $v$, number of items $n$
        \State $\mathcal{S} \gets \emptyset$
        \State $\mathcal{V} \gets \emptyset$
        \For{$i = 0$ up to $B + 1$} \State $\mathcal{V}[i] \gets 0$ \Comment{initializing DP array} \EndFor
		\For{$i = 1$ up to $n + 1$}
            \For{$j = B$ down to $0$}
            \State \textbf{if} $w[j - 1] \leq j$ and $\mathcal{V}[j - w[i - 1]] + v[i - 1] > \mathcal{V}[j]$
            \Indent \State $\mathcal{S} \gets \mathcal{S} \cup (j-1)$ \EndIndent
            \State \textbf{else} 
            \Indent \State $\mathcal{S} \gets \mathcal{S} \setminus (j-1)$ \EndIndent
		    \State $\mathcal{V}[j] \gets \textbf{max} \left( \mathcal{V}[j], \mathcal{V}[j - w[i - 1]] + v[i - 1] \right)$  
        \EndFor
        \EndFor
        \State \textbf{return} $\mathcal{S}$ \Comment{return set of selected items}
	\end{algorithmic}
\end{algorithm}	
\setlength{\textfloatsep}{\textfloatsepsave}



\subsection{Equitable Allocation Extension}

In the \texttt{E-NHPT} problem, we seek to select candidates subject to $k$ group-specific budget constraints, which define an equitable partitioning of the total budget.  Since this seeks to constrain the \textit{investment} allocated to any given group, the problem is an instance of group-fair knapsack where fairness is measured according to the \emph{weight} of items accepted from each group.

Prior work has shown that this group-fair setting of knapsack can be cast as a special case of the multidimensional knapsack problem \cite{PatelKhanLouis:2021}.  Therefore, we employ a partitioning technique to split the candidate selection problem into $k$ separate knapsack problems, where $k$ is the number of groups under consideration for the allocation.
Given each subset $\mathcal{N}_z$, representing the neighborhoods which belong to group $z$, we construct the $z$th instance of the knapsack problem -- items are neighborhoods in $\mathcal{N}_z$, with a corresponding list of values (carbon reduction) and weights (cost).  The knapsack's capacity is $b_z$, where $b_z$ is the maximum allocation for the $z$th group. We denote the selected item set by $\mathcal{S}_z$.

With $k$ separate solutions to the knapsack problem of the form $\mathcal{S}_z$, the solution to the group-fair knapsack problem is constructed by taking the union of all group-specific item sets: $\mathcal{S} = \bigcup_{z = 1}^{k} \mathcal{S}_z$.

\section{Experimental Setup}
\label{sec:expmethods}

In this section, we describe our experimental setup including gas \& electric usage data sets, census \& property data, gas distribution network details, and cost models. 


\vspace{-0.2cm}
\subsection{Data sets}
\noindent\textbf{Gas \& electric usage data.}
We use gas and electric usage data collected from a small city in the northeastern United States.
The gas and electric network in this city is operated by a municipal utility and is a representative of other networks in similar climates. 
Thus, the insights presented in this work are widely applicable.

Table~\ref{tab:charecteristics} provides the summary of the data set. Gas usage and electricity usage data is collected at one hour and five minute granularity using 6,445 gas meters and 13,800 electric meters, respectively. 
The metadata provides a mapping of both gas and electric meters to buildings in the city and a mapping of electric meters to distribution transformers.  
We use this data to generate usage profiles for homes and peak load for each distribution (edge) transformer in the electric grid. 
We use data for the entire 2020 calendar year as it provides a complete snapshot of usage across seasons.  

\begin{table}[]
\caption{Summary of data set characteristics}\label{tab:charecteristics}
\vspace{-0.35cm}
\begin{tabular}{|l|l|l|}
\hline
                 & \textbf{Number of Meters} & \textbf{Time Interval} \\ \hline
Electric Data    & 13,800         & 5 minutes     \\ \hline
Natural Gas Data & 6,445          & 60 minutes    \\ \hline\hline\hline
\textbf{Duration}         & \multicolumn{2}{l|}{1/1/2020 - 12/31/2020} \\ \hline\hline
\end{tabular}
\vspace{-0.2cm}
\end{table}

\smallskip
\noindent\textbf{Property \& census data.}
For buildings in the usage data set, we collect information such as building type and zoning from publicly available tax records containing property data.  
We map the addresses in the property data set to the buildings in the usage data set.
Since we focus on residential households, we exclude commercial and industrial buildings from analysis. 

We also gather data reported by the U.S. Census Bureau~\cite{Census:2020} to identify socioeconomic characteristics of the households and neighborhoods in our data set.  
We classify households into low, medium, and high income neighborhoods using the median income, an approach inspired by prior work on equitable transition~\cite{WamburuGrazierIrwinCragoShenoy:2022}.






\subsection{Gas Network Approximation}\label{sec:networkapprox}

Although our usage data set includes gas meter locations, the exact topology of the city's gas mains and service pipelines is not provided. 
To facilitate a network-aware transition strategy, we need a reasonable estimate of the gas distribution network in the city. 
Prior work states that natural gas distribution lines are predominantly buried underneath vehicle roads~\cite{Jackson:2014}.  
We use field knowledge and insights from prior work to estimate the network topology. 

\begin{figure}[t]
    \centering
    \begin{tabular}{cc}
    \includegraphics[width=0.2\textwidth]{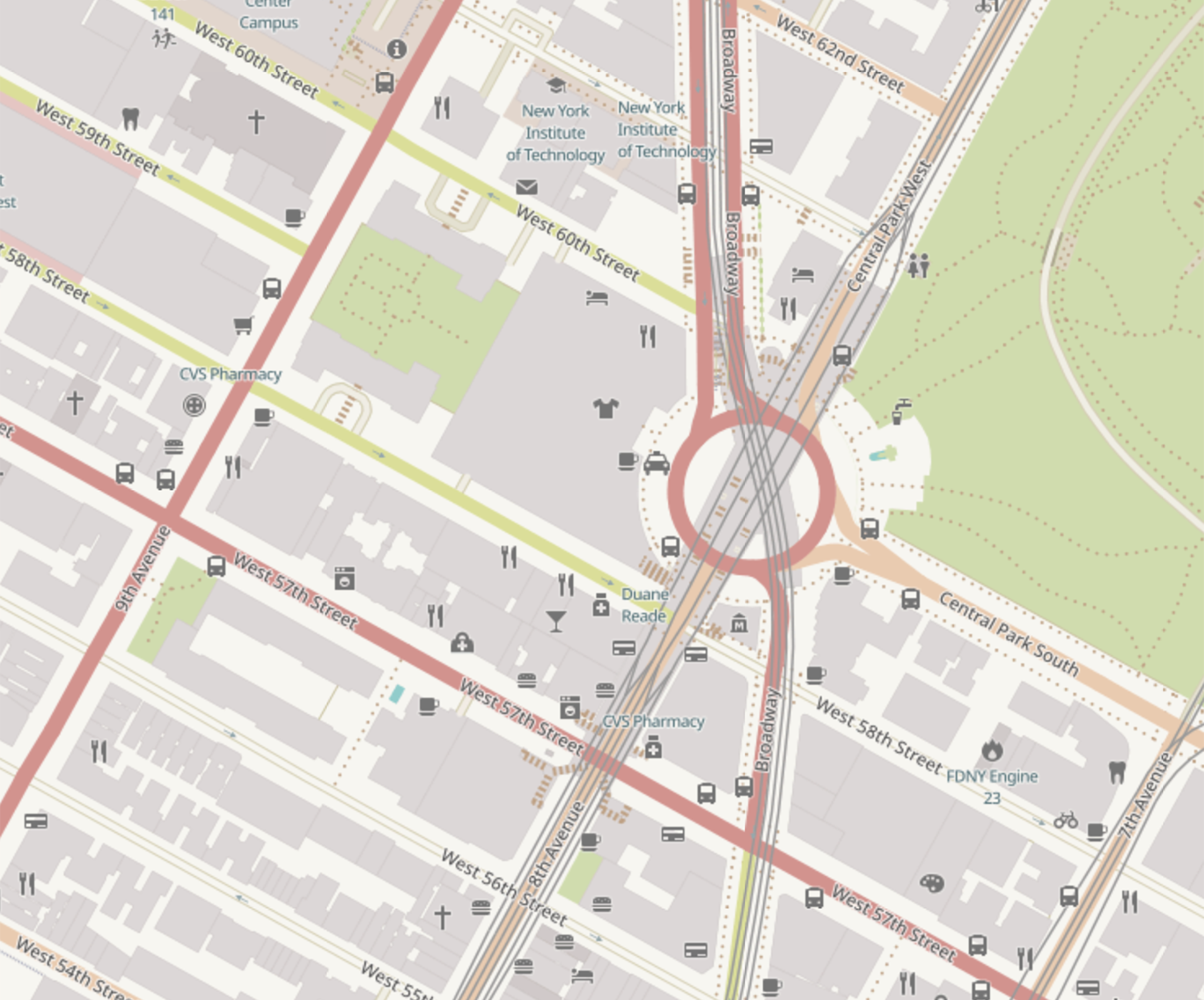} &
    \includegraphics[width=0.2\textwidth]{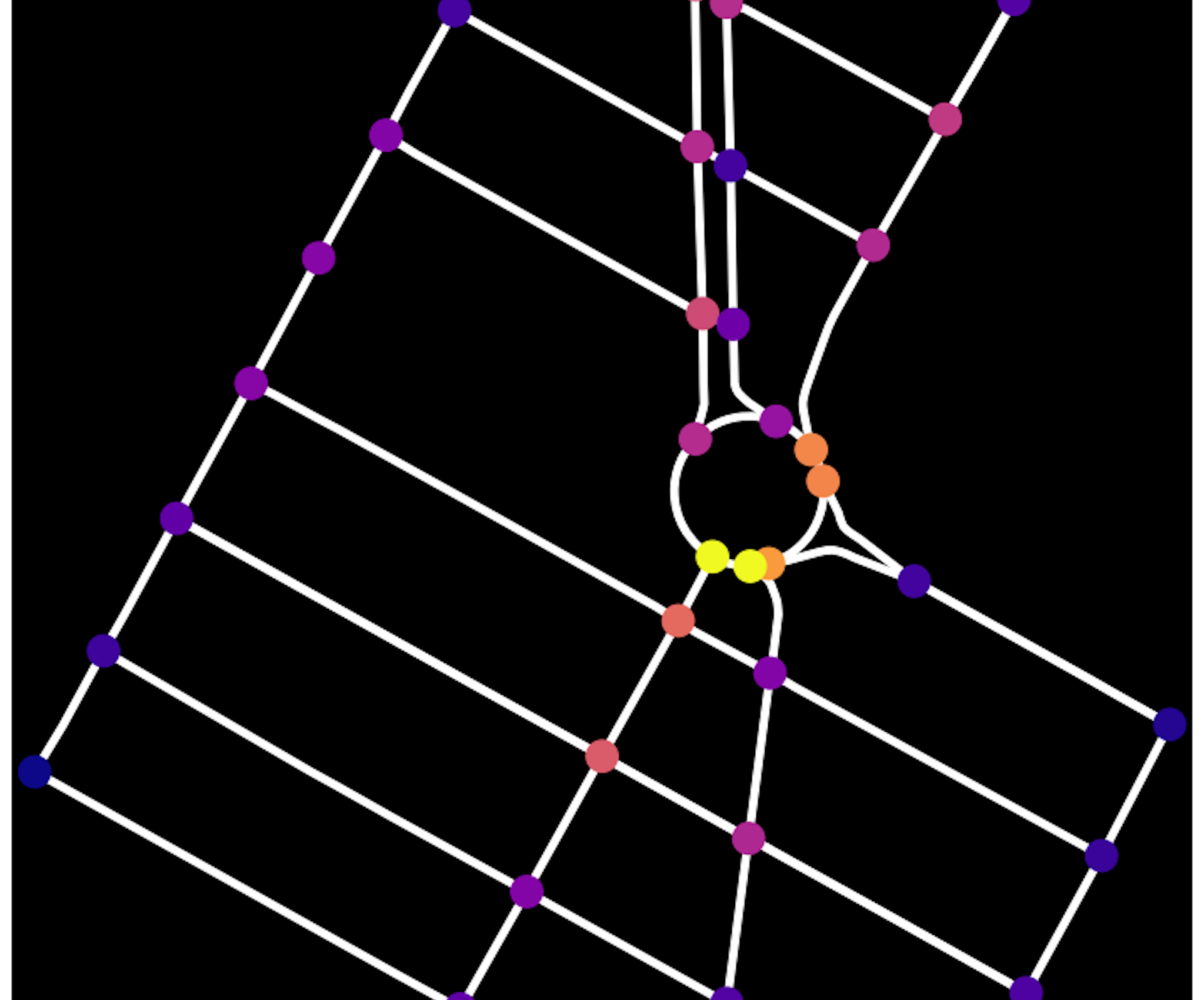}\\
    (a) OpenStreetMap data&
    (c) NetworkX graph\\
    \end{tabular}
    \caption{OpenStreetMap data (a) and its representation as a NetworkX graph (b) for an example city segment.}
    \label{fig:mapExample}
    \vspace{-0.7cm}
\end{figure}

We use road network data from OpenStreetMap for the entire city~\cite{OpenStreetMap}.
We process the data and convert it into a NetworkX graph representation using the OSMnx Python package~\cite{HagbergSchultSwart:2008, Boeing:2017}.  
In this graph, each edge represents the geometry and location of a road segment, while nodes represent any type of junction between roads.
We leverage field knowledge of our utility partners to filter the network further -- for example, an interstate is unlikely to cover parts of the local gas distribution network, while a main thoroughfare in the city center is likely to be co-located with the distribution pipeline.
Figure~\ref{fig:mapExample} depicts the OSM data and its representation as a NetworkX graph for several blocks of an example city.

\noindent\textbf{Flow modeling \& network simplification.}
In a gas distribution network, supply flows from a transmission pipeline through a \emph{gate station}, which regulates the pressure and serves as the single point of entry into the distribution network.  
From a graph theoretical perspective, the gate station can be considered the \emph{source}.
We know the location of the gate station for this case study, and use it as a source node to simplify the OSMnx graph. 

We use Dijkstra's single-source shortest path algorithm to compute the shortest paths from the source to all other nodes in the network.  
We delete edges which are not used by any shortest path, removing $\sim$68\% of the graph's edges and leaving a directed graph which approximates gas flow outwards from the source node to each point in the city.
We also delete any nodes disconnected from the main network, which represent anomalous data points from OpenStreetMap.  This ensures that the network is weakly connected.

\noindent\textbf{Meter connections in the approximate network.}
We assign households in the usage data set to appropriate segments of gas distribution network. 
To do so, we use each household's location to assign that household's gas meter ID to the closest edge.
We also store the distance from the edge to the household location, which acts as a proxy for the pipe length required to tap into the gas main.






\smallskip

\noindent\textbf{Defining candidate neighborhoods for transition.}
We next divide the approximate gas network into \emph{neighborhoods} that serve as a unit for transition in our framework. 
We begin with an arbitrary edge $(u,v)$ from the graph, which is the smallest unit that allows a full gas shutdown. In an OSMnx graph, an edge roughly represents ``one block''. 
To simulate a shutdown, we delete edge $(u,v)$, convert gas usage of households assigned to that edge to equivalent electric load, and delete nodes $u$ or $v$ if either has degree~$= 0$.  The length of the edge and lengths of all pipelines connecting households to the gas main are saved to calculate future maintenance savings.

If the graph $G$ remains weakly connected after the deletion of edge $(u,v)$, we say that edge $(u,v)$ is a ``psuedo-index'' for the neighborhood $N_{(u,v)}$, which contains one edge $\{(u,v)\}$.
If the graph $G$ is \emph{no longer} weakly connected after the deletion of edge $(u,v)$, $G$ contains two connected components $C_1$ and $C_2$.  
Without loss of generality, let $C_1$ be the component which contains the source node for gas network.  
Since $C_2$ is no longer connected to the source node, shutting down gas service at edge $(u,v)$ prevents gas flow to component $C_2$.  
This suggests that we must also convert all of $C_2$.  
Let $\mathcal{E}_2$ denote the edges in $C_2$.  
We state that the edge $(u,v)$ is a ``pseudo-index'' for the neighborhood $N_{(u,v)} = \{(u,v)\} \cup \mathcal{E}_2$.  
Each edge in $\mathcal{E}_2$ is iteratively deleted from the graph and households are ``transitioned'' to heat pumps using the procedure described above.

We repeat the process for each edge $(i,j) \in \mathcal{E}$ to define a set of neighborhoods $\mathcal{N}$ considered for transition in our framework.

\noindent\textbf{Estimating carbon emissions.}
To calculate \emph{carbon emissions} of a household's gas usage, and by extension the carbon intensity of each neighborhood or the entire network, we convert gas usage into the equivalent amount of CO$_2$.  For each hundred cubic feet (CCF) of natural gas used, $0.00551$ metric tons of CO$_2$ are emitted~\cite{EPA:2018}.

\noindent\textbf{Converting gas demand to electric demand.}
We next compute the amount of electrical energy required for an ASHP to move the equivalent heat energy generated by the recorded gas usage.
We first compute the heat energy in British Thermal Units (BTU) from gas usage data assuming a gas furnace efficiency of 87.5\%, which is the average of efficiencies for a standard and an efficient unit~\cite{Brand:2012}. 

To compute electrical energy in kilowatt-hours (kWh), we assume that the ASHP has an average Coefficient of Performance (COP) of 2.5, which is easily attainable by modern systems \cite{KellyCockroft:2011}.  The electric usage in kWH can then be computed using this formula: $\textnormal{Electric Usage in kWh} = \left( \frac{\texttt{Heat Energy}}{COP} \cdot \frac{0.293071}{10^{3}} \times \frac{\textnormal{kWh}}{\textnormal{BTU}} \right)$.
This additional demand is added to the existing load on the distribution transformer supplying electricity to the building. 

\noindent\textbf{Defining groups for equitable transition.}
Prior work suggests that a neighborhood's median income is an effective predictor of many relevant socioeconomic quantities~\cite{WamburuGrazierIrwinCragoShenoy:2022}.
Based on this, we leverage census data to create income profiles of neighborhoods in the city.  
For simplicity, we define three \emph{income groups}: low, medium, and high.  
We categorize each neighborhood into these groups based on median household income.
Cutoffs for the groups are based on split points which roughly divide the city into thirds; the low income group has annual median income below $\$45,000$ USD, the high income group has annual median income above $\$80,000$ USD, and the middle income group contains all other neighborhoods.


\vspace{-0.25cm}
\subsection{Cost Modeling}
\label{sec:cost-models}
In this section, we describe empirical cost models for different components of the budget constraint of our framework.


\noindent\textbf{Gas distribution network maintenance cost.}
Utilities spend a significant portion of their budget to maintain and upgrade the gas distribution network.
They can save money in perpetuity if portions of the network are permanently shut down.  The cost of maintaining a gas network depends on various factors including length, age, material, and diameter of pipe. 
Utilities may prioritize decommissioning certain segments of the gas network based on their age or material -- older pipelines have low stranded cost and help avoid pipeline upgrades, while some materials such as cast iron are more leak-prone than others~\cite{Gallagher:2015}. 
However, such information is not available and cannot be inferred from publicly available data.

To approximate maintenance costs, we use the total length of the gas network and the average yearly maintenance expenses for the city, which are publicly available~\cite{holyoke-docs}.  
We use this information to compute per unit length cost of maintaining the network for one year. 
Then, if X\% of the total pipeline length is decommissioned, we count X\% of the maintenance costs as savings.
This heuristic is motivated by prior work that shows the cost to maintain underground pipelines grows linearly with the total length of the pipelines~\cite{Najibi:2009}.
Measuring the yearly maintenance savings \emph{does not} fully capture the potential savings that the utility can realize by eliminating maintenance in perpetuity, but serves as a useful proxy.

\smallskip
\noindent\textbf{Transformer upgrade cost.}
The electric grid and especially edge transformers must accommodate the additional load of ASHPs.
As discussed in~\S\ref{sec:background}, limiting the amount and duration of overloading events is an important consideration to maintain transformer lifetimes and reduce grid outages.



\begin{figure}[t]
    \includegraphics[width=0.85\linewidth]{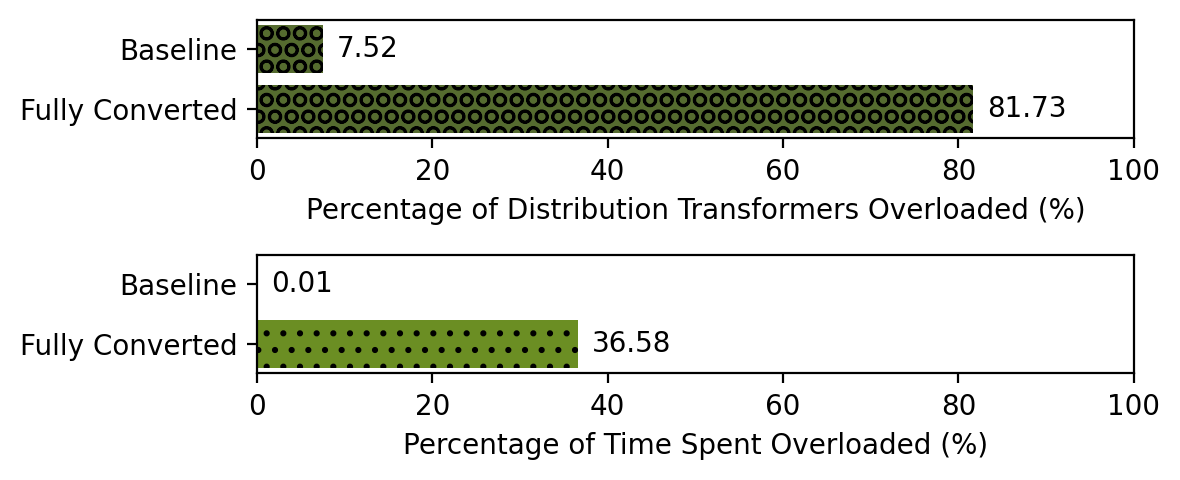}
    \vspace{-0.5cm}
    \caption{Percentage of the overloaded transformers (top) and percentage of time spent in overload for transformers (bottom) under the current scenario (baseline) and under 100\% transition to electric heat pumps (fully converted).}
    \label{fig:fullASHP}
    \vspace{-0.5cm}
\end{figure}

For our analysis, we consider a transformer with a peak load between 90\% and 125\% of its capacity to be \textit{highly utilized}, and a transformer with a peak load above 125\% of its capacity to be \textit{overloaded}.
Any transformer that becomes overloaded post-transition would need to be upgraded. 
Figure~\ref{fig:fullASHP} shows the percentage of overloaded transformers and the time they spend overloaded at present and under a hypothetical 100\% heat pump transition scenario. 
We observe that a full transition to electric heat pumps would require more than 80\% of transformers to be upgraded. 
Therefore, utilities want to avoid widespread upgrades in favor of a targeted transition that defers upgrades as far into the future as possible. 



In our study, we consider the cost of upgrading transformers and other components such as distribution and feeder lines. 
Prior work has estimated the cost of grid upgrades in other scenarios~\cite{PaloAlto:2020, Sahoo:2019,  Brinkel:2020}. 
We use these studies to devise a rule-based model that estimates the cost of distribution grid upgrades with parameters of overloaded transformers as an input.  
Since we do not have information on distribution and/or feeder lines, we leverage a heuristic from prior work to estimate the total cost of upgrading non-transformer components of the electric grid including labor~\cite{PaloAlto:2020}.

\vspace{-0.15cm}
\begin{definition}[Overloaded Transformer Upgrade Rules]\label{def:trans}
Let overloaded transformer $i$ have existing capacity $c(i)$ and overloaded peak load $L^{\texttt{peak}}(i) > 1.25 c(i)$.

If $c(i) \leq 75$ kVA, we assume that $i$ is a pole-top transformer.

If $c(i) > 75$ kVA, we assume that $i$ is a pad-mount transformer.

For pole-top transformers, transformers are available in [15, 25, 37.5, 50, 75] denominations and have a cost in \$4,225 - \$25,525 range. 
For pad-mount transformers, transformers are available in [75, 100, 150, 167] denominations and have a cost in \$74,900 - \$149,800 range. 

\begin{itemize}
    \setlength\itemsep{-0.3cm}
    \item \textbf{if} $L^{\texttt{peak}}(i) < 2 \cdot c(i)$ and $L^{\texttt{peak}}(i) \leq 75 \rightarrow $  Upgrade existing pole-top transformer to larger capacity pole-top transformer.\\
    \item \textbf{if} $L^{\texttt{peak}}(i) > 2 \cdot c(i)$ and $c(i) \leq 75 \rightarrow $  Install additional pole-top transformer to augment grid capacity.\\
    \item \textbf{if} $L^{\texttt{peak}}(i) > 3 \cdot c(i)$ and $L^{\texttt{peak}}(i) > 75 \rightarrow $  Replace existing pole-top transformer with new pad-mount installation.\\
    \item \textbf{if} $c(i) > 75 \rightarrow $  Install additional pad-mount transformer to augment grid capacity.\\
\end{itemize}
\vspace{-0.45cm}
When a transformer is upgraded, we assume that the capacity purchased is the nearest denomination which satisfies the required load.
\end{definition}
\vspace{-0.25cm}





This cost model does not include context such as the existing transformer age or information about which transformers are scheduled for imminent replacement. 
We can incorporate cost estimates with such information in the future.

\begin{figure*}[t]
	\minipage{0.23\textwidth}
	\frame{\includegraphics[width=\linewidth]{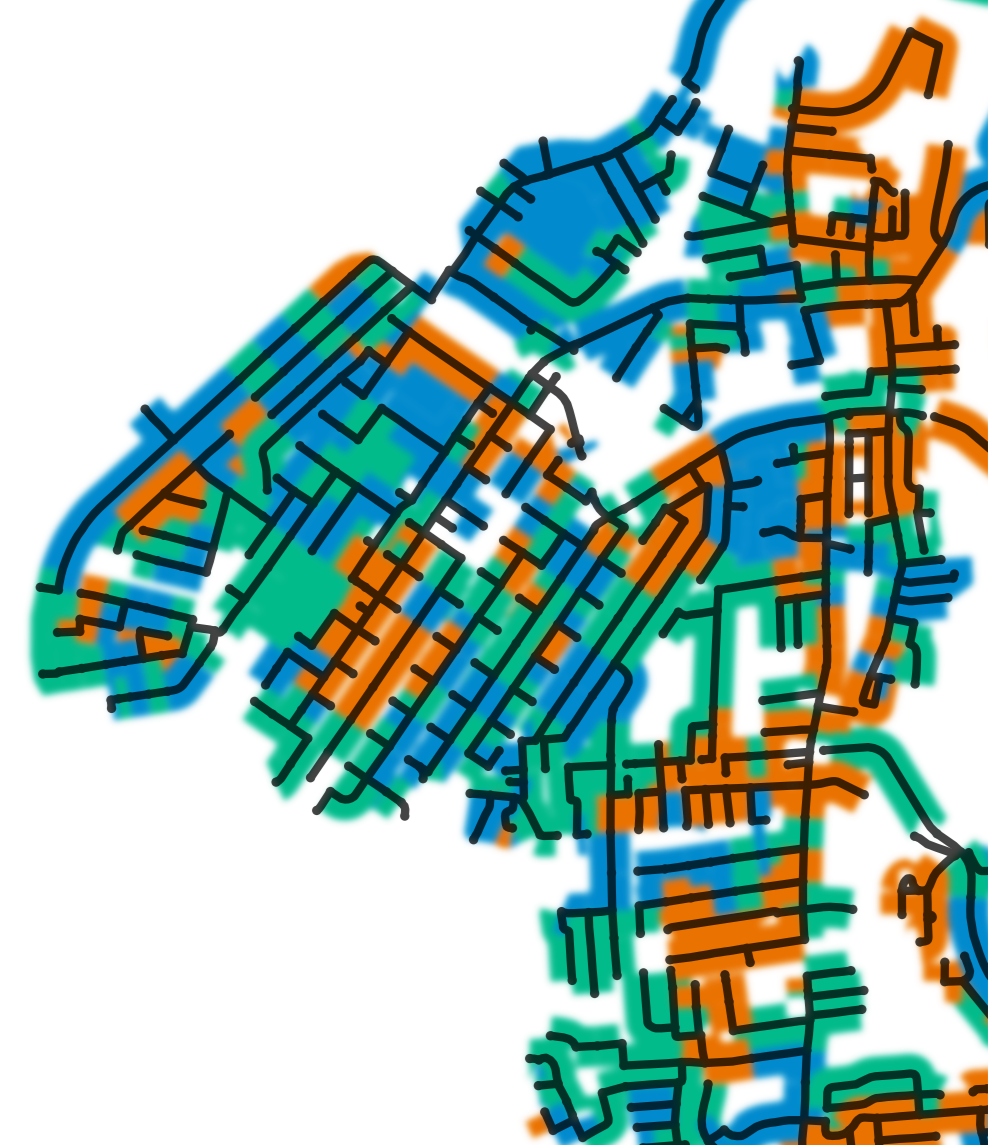}}\vspace{-1em}
    \caption*{(a) Initial gas network}
	\endminipage\hfill
	\minipage{0.222\textwidth}
	\frame{\includegraphics[width=\linewidth]{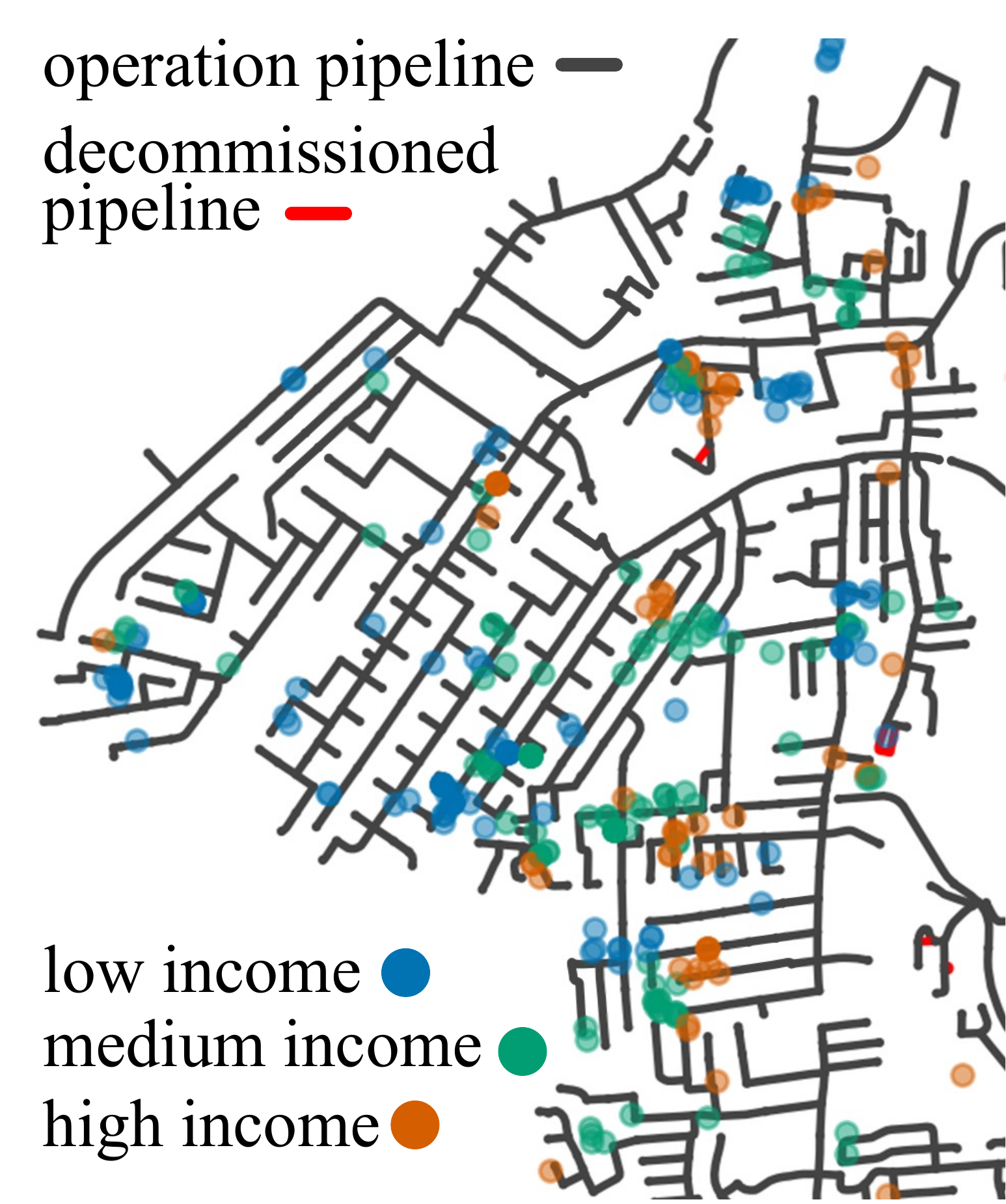}}\vspace{-1em}
    \caption*{(b) Network-oblivious} 
	\endminipage\hfill
    \minipage{0.23\textwidth}
	\frame{\includegraphics[width=\linewidth]{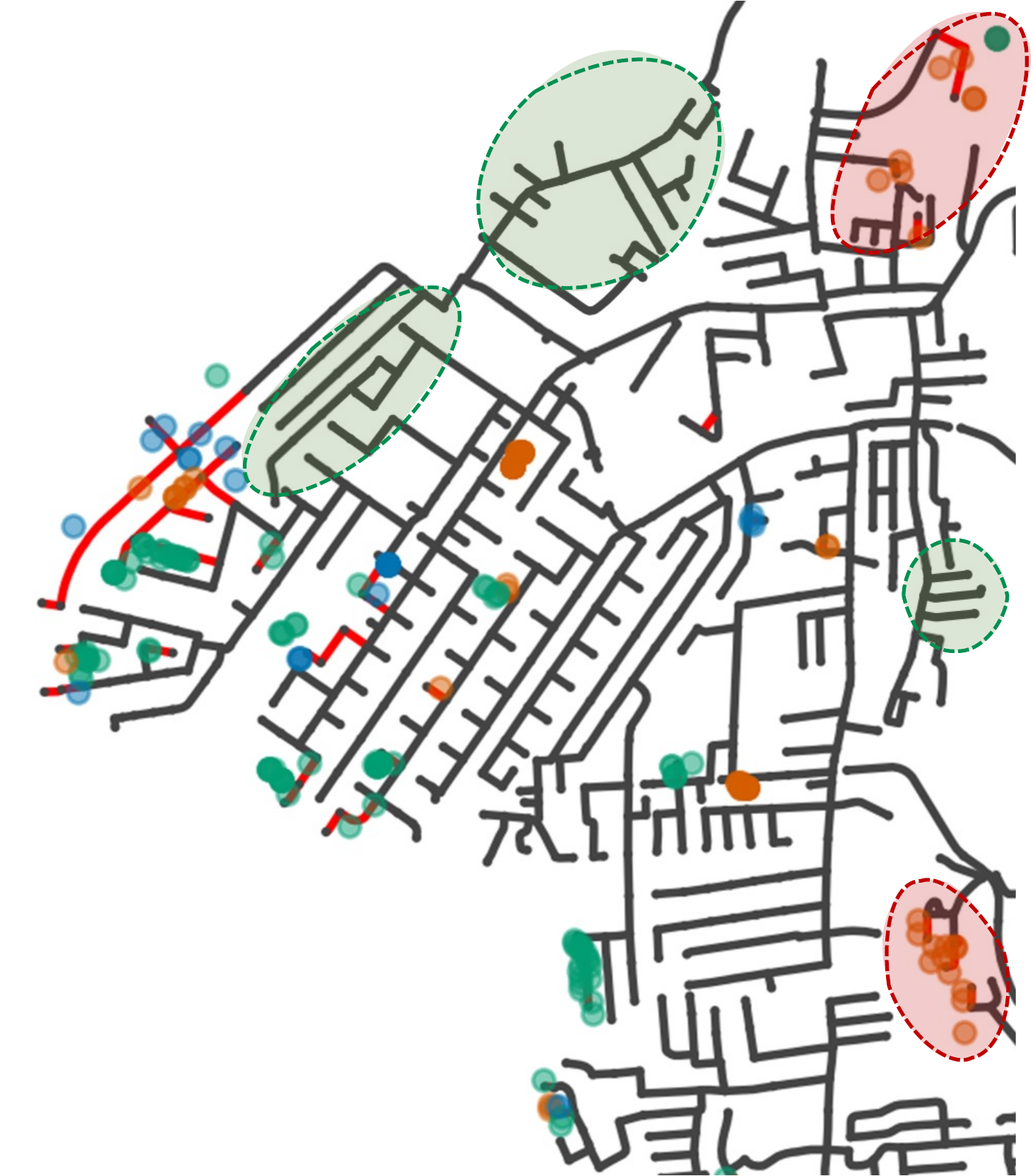}}\vspace{-1em}
    \caption*{(c) Network-aware} 
	\endminipage\hfill
    \minipage{0.23\textwidth}
	\frame{\includegraphics[width=\linewidth]{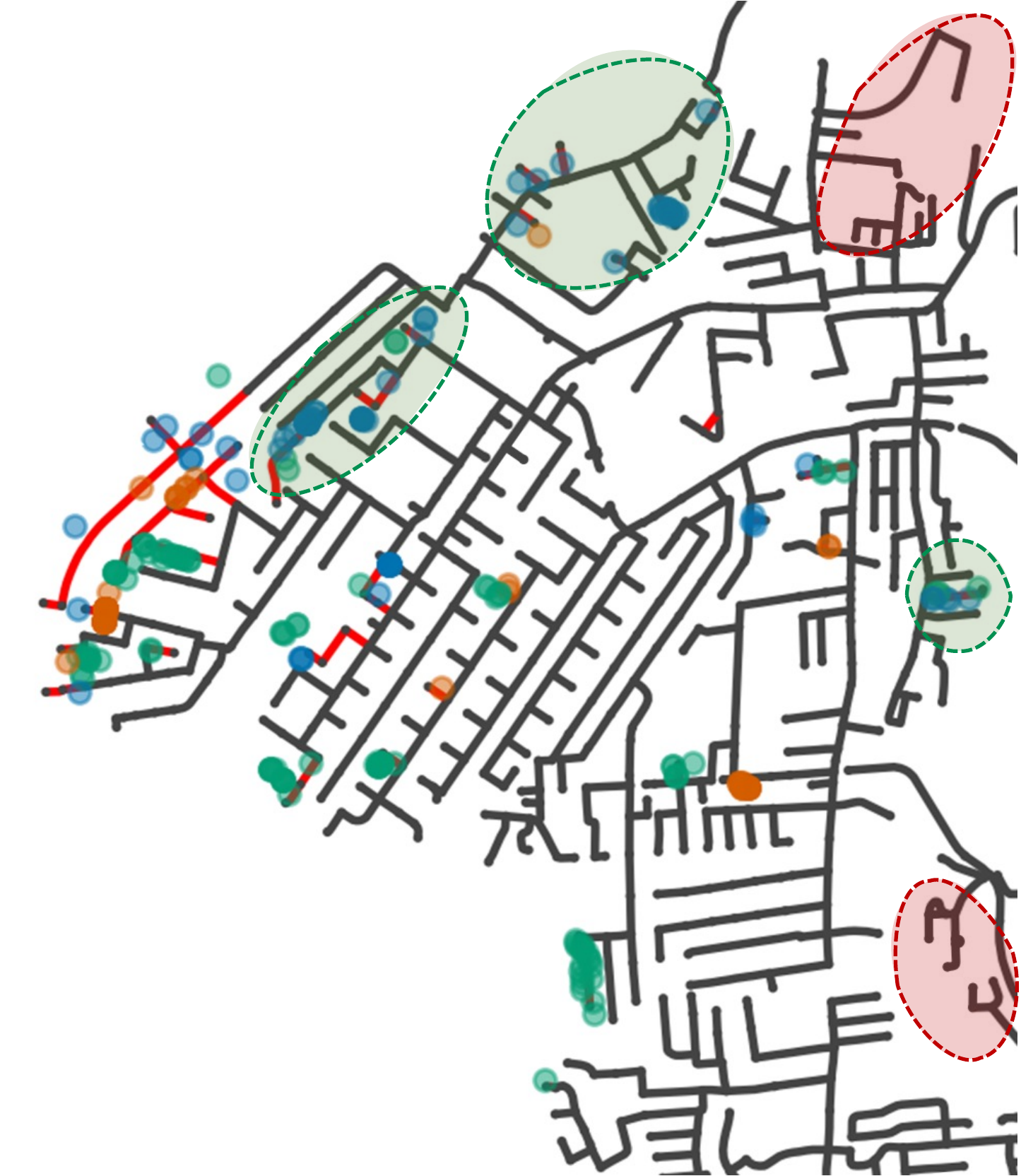}}\vspace{-1em}
    \caption*{(d) Equity-aware} 
	\endminipage\hfill
	\vspace{-0.4cm}
    \caption{A representative snapshot of city's initial gas network (a) and final gas networks with marked decommissioned pipelines and transitioned households under \emph{network-oblivious} (b), \emph{network-aware} (c), and \emph{equity-aware} (d) transition strategies. }
    \label{fig:visualResults}
\end{figure*}

\smallskip
\noindent\textbf{Electric heat pump installation cost.}
The upfront cost of EHPs can be prohibitive, especially for low-income and marginalized populations which will experience a disproportionate burden of climate change~\cite{Wilson:2020, Salas:2021, Gutschow:2021}. 
As a result, governments and utilities worldwide have started incentive programs to provide heat pump installation rebates~\cite{Snape:2015, MassSave, NRC:2021, CENHUD}.
In our study, we assume that utilities bear a significant portion of the upfront installation costs to provide an attractive incentive for communities to transition.  

While we do not have data to support a percentage of households which would be willing to accept this transition, prior work \cite{WamburuLeeShenoyIrwin:2018} has shown that EHP installation yields significant economic benefit, reducing a household's heating costs by up to 60\%.



We size the ASHP system for each household based on the median gas usage for heating. 
To set a benchmark system, we use a household with the median gas usage and estimate the \emph{transition cost} -- including equipment, installation, and labor -- to be \$15,000 USD for a high-efficiency residential ASHP system~\cite{Desai:2022}.  
We normalize the cost of the system with the median usage to estimate the cost for a building with different gas usage. 
As an example, a building with $2\times$ the median usage will incur a transition cost of \$30,000 USD.
Additional household information such as area and insulation can improve the accuracy and granularity of the model. 







\section{Experimental Evaluation}
\label{sec:eval}
In this section, we present an experimental evaluation of our optimization framework using data from a city in the Northeast U.S. 

\noindent\textbf{Transition strategies.}
We evaluate the performance of our \emph{network-aware} and  \emph{equity-aware} transition strategies against a baseline \emph{network-oblivious} strategy from prior work~\cite{WamburuGrazierIrwinCragoShenoy:2022}. 
The \emph{network-oblivious} approach transitions households that yield the highest carbon emission reductions without considering the impact on the gas network and subsequent electric grid upgrade costs. 

\noindent\textbf{Metrics.}
We compare the strategies across following key metrics: carbon reduction achieved, number of households transitioned, extent of the gas network shutdown, number of transformers upgraded, and distribution of transitioned households across income groups (equity-specific metric). 
All values are reported in percentage.
A higher value is good for all but the last two metrics.
A strategy should require upgrading a small number of transformers and select proportionate number of households across all income groups. 

\noindent\textbf{Parameters.} 
In our analysis, we vary the total budget as a percentage of a \emph{benchmark budget}. 
The \emph{benchmark budget} includes the cost of ASHP installation in all homes currently using gas heating, and all relevant distribution grid upgrades.
Since our estimates for gas network maintenance cost are approximate, we also analyze the effect of varying maintenance cost in our metrics. 
We use the gas network maintenance cost computed in \S~\ref{sec:cost-models} as a \emph{reference cost}.

\noindent\textbf{Relevant data statistics.} 
Our evaluation of transition strategies comments on the gas usage and spatial locality of homes in various income tracts. 
To support such statements, we present high level statistics from our dataset. 
First, households in high- and medium-income tracts tend to be bigger and have higher gas usage. 
In our data set, medium- and high-income households have 27.29\% and 46.99\% higher gas usage than low-income households, respectively. 
Second, medium- and high-income households tend to be more sparsely located than low-income households. 
In our data set, we have 9.66, 10.89, and 13.21 average homes per block for high, medium, and low-income households, respectively. 
An example insight from the data is that high-income neighborhoods are an attractive option for gas network shutdown (less number of heat pumps needed per pipeline shutdown) but the benefits are reduced due to higher cost of heat pumps (larger size of heat pumps due to high gas usage).

\vspace{-0.1cm}
\subsection{A Bird-eye View of Transition Strategies}

In Figure~\ref{fig:visualResults}, we show a representative snapshot of the city's gas network before and after various transition strategies.
In Fig.~\ref{fig:visualResults}a, we shade neighborhoods based on the median income to show the wide disparity in income-levels that exists across the city (labels in \ref{fig:visualResults}(b)). 
We observe that the \emph{network-oblivious} strategy transitions high emitting households spread across the city. Our \emph{network-aware} strategy transitions households situated close to each other to reduce network costs. Finally, our \emph{equity-aware} strategy picks low- and middle-income households (green shaded areas) in place of high-income households (red shaded areas) to ensure equitable allocations. 
We further discuss, and refer back to, this figure as we evaluate our strategies in subsequent sections. 



\begin{figure}[t]
    \includegraphics[width=\linewidth]{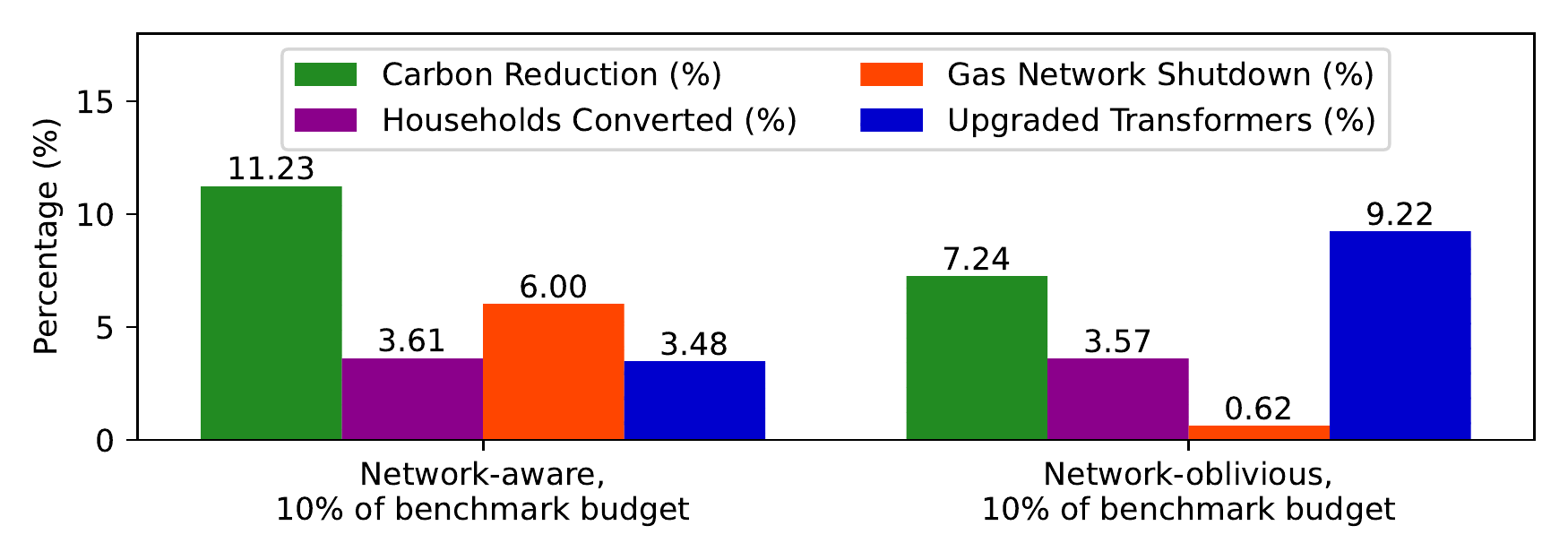}
    \vspace{-0.9cm}
    \caption{Our \emph{network-aware} strategy (left) outperforms the \emph{network-oblivious} strategy (right) across all the metrics at 10\% of the benchmark budget.}
    \label{fig:naiveVSnetwork}
    \vspace{-0.75cm}
\end{figure}

\begin{figure*}[t]
    \centering
    \begin{tabular}{cc}
    \includegraphics[width=0.46\textwidth]{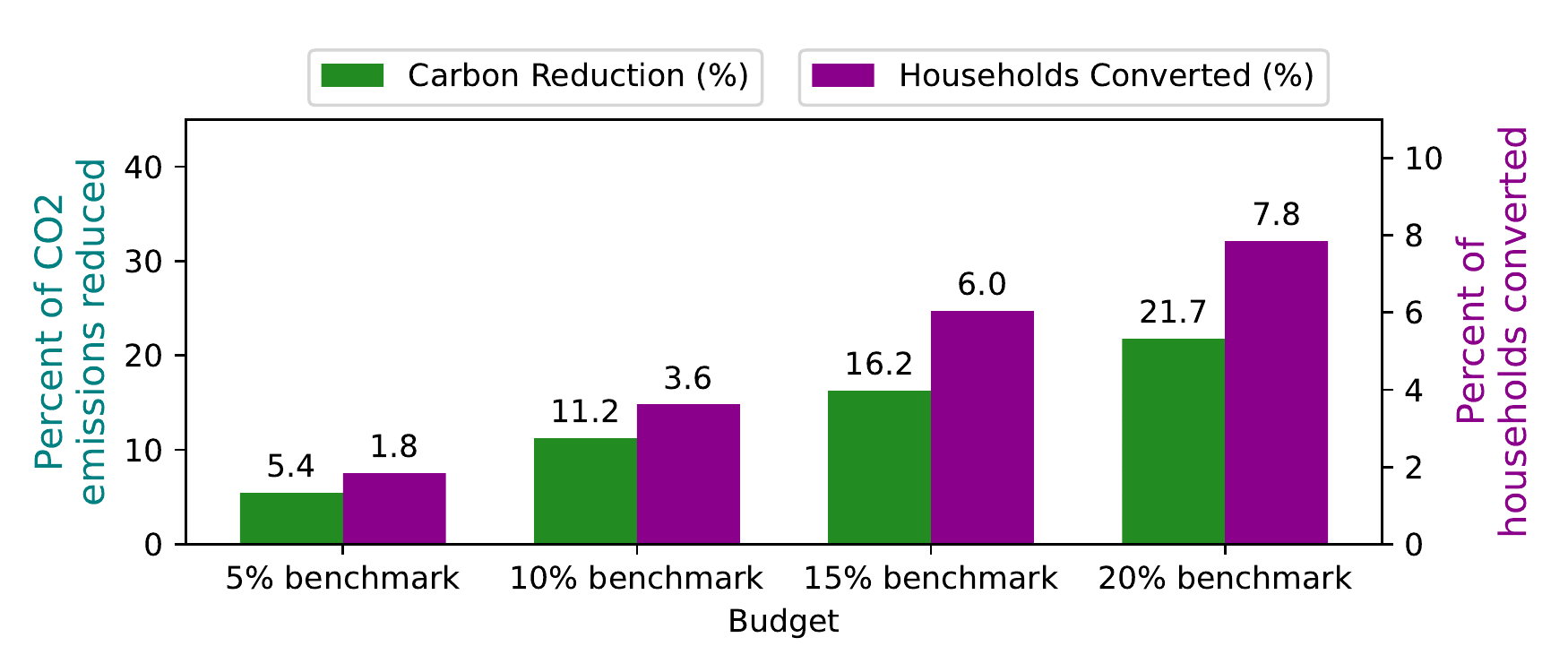} &
    \includegraphics[width=0.46\textwidth]{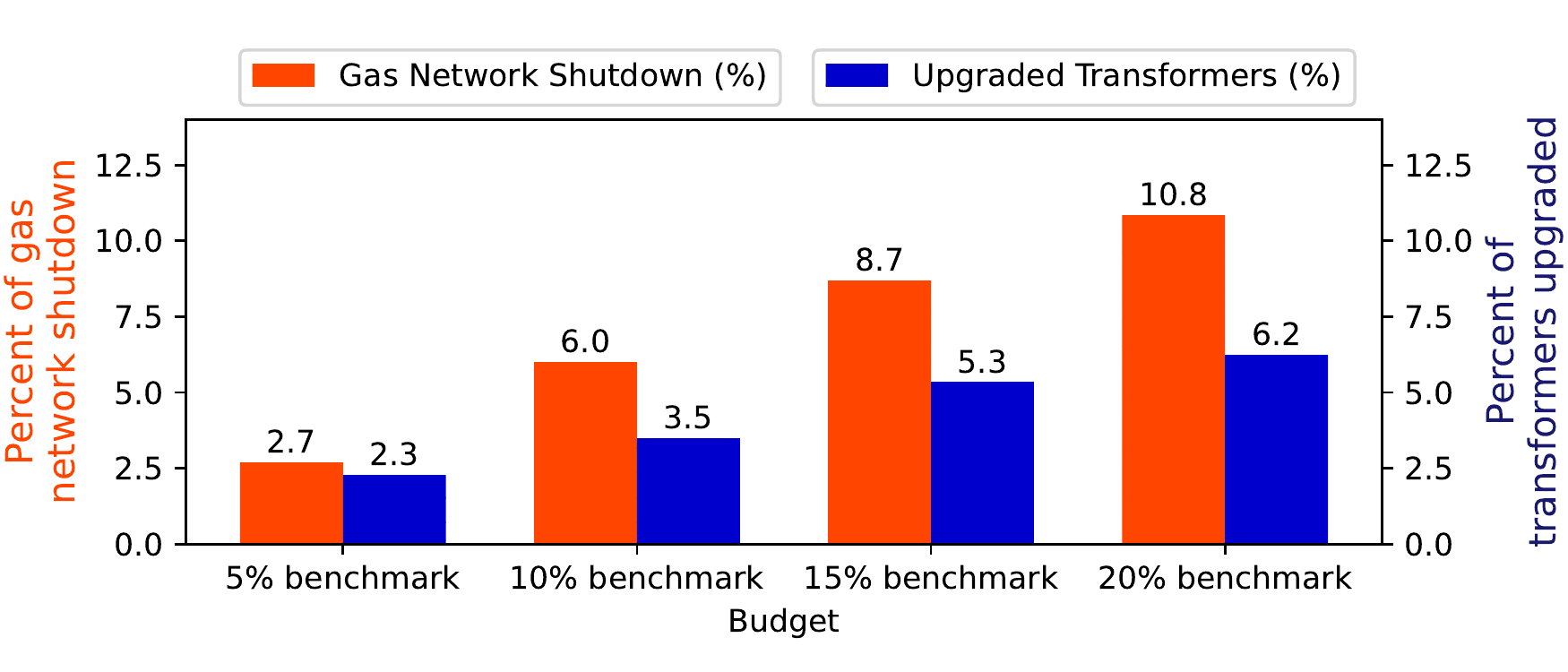}\vspace{-0.2cm}\\
    
    (a) Carbon Reduction \& ASHP Conversions &
    (b) Gas Network Shutdowns \& Grid Upgrades\\
    \end{tabular}
    \vspace{-0.4cm}
    \caption{The performance of \emph{network-aware} strategy at varying percentages of benchmark budget.}
    \label{fig:results1}
    \vspace{-0.3cm}
\end{figure*}

\begin{figure*}[t]
    \centering
    \begin{tabular}{cc}
    \includegraphics[width=0.46\textwidth]{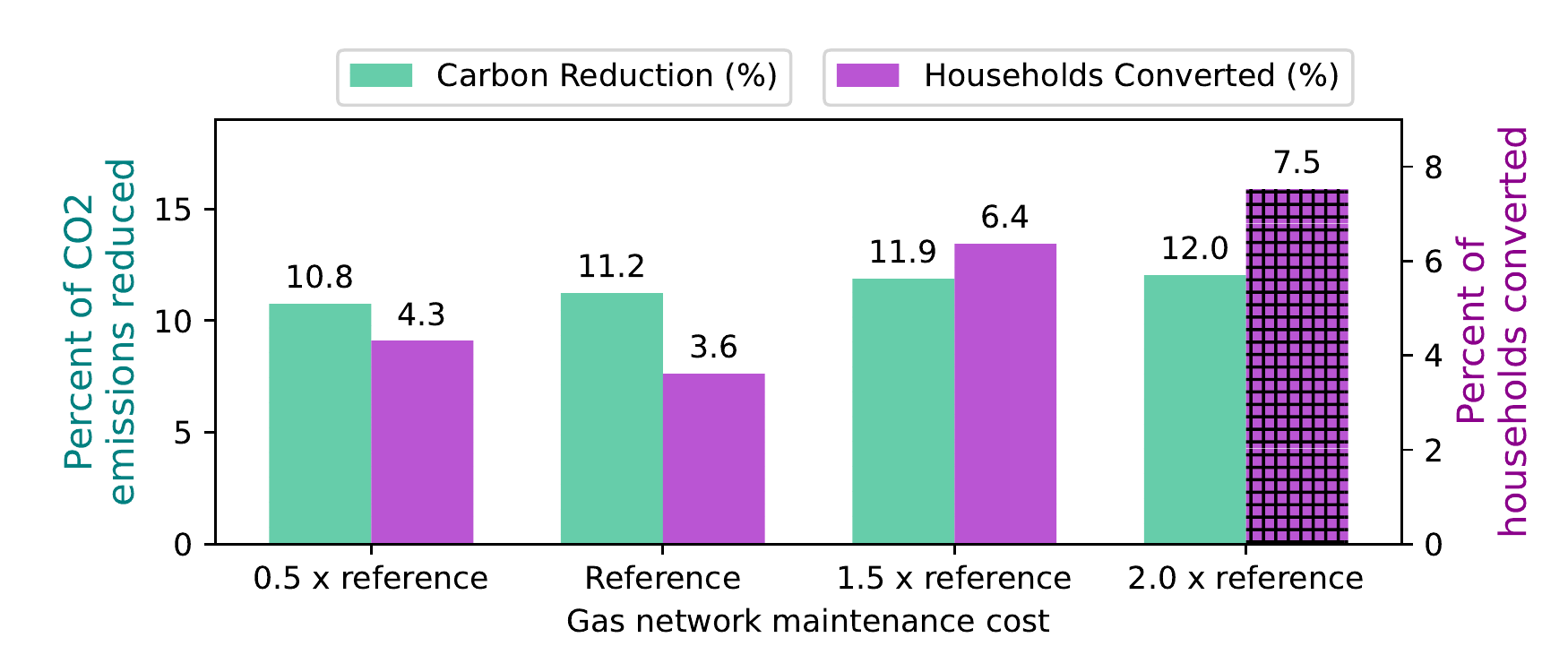} &
    \includegraphics[width=0.46\textwidth]{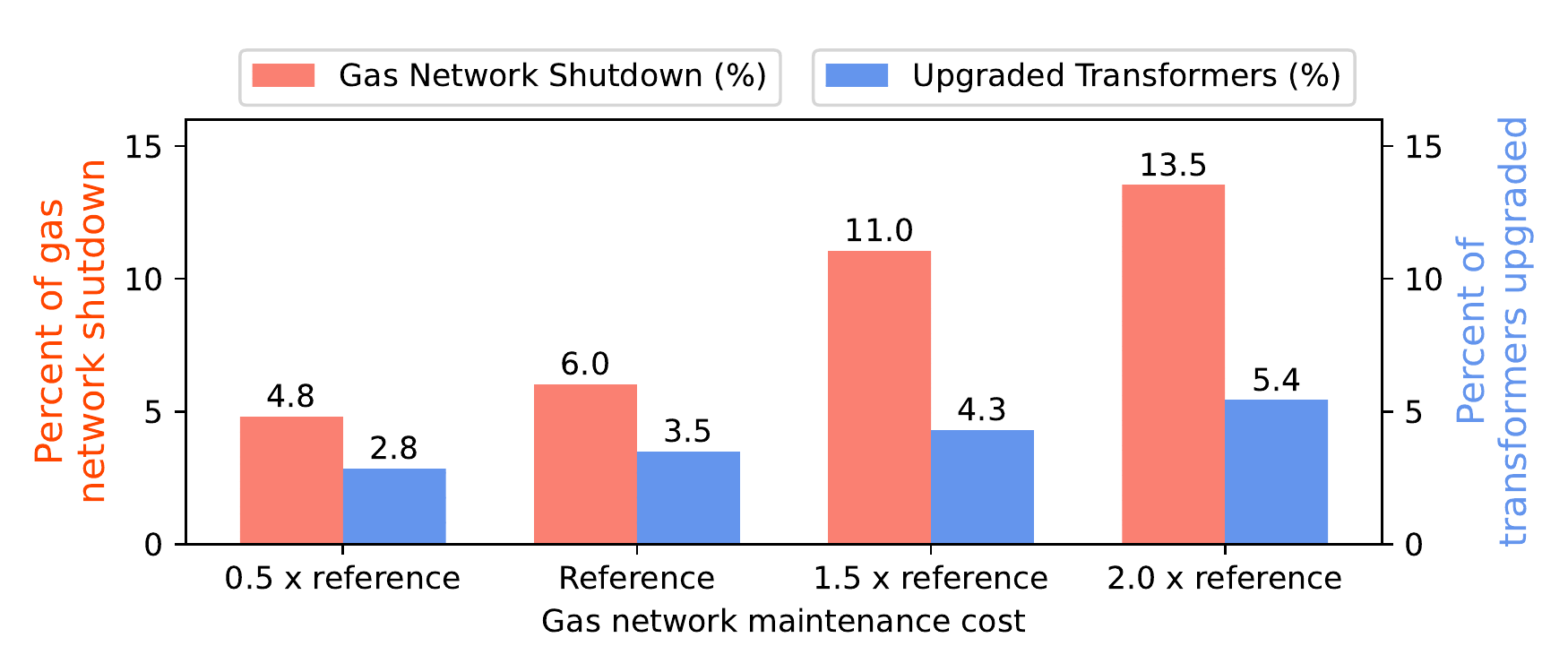}\vspace{-0.2cm}\\
    (a) Carbon Reduction \& ASHP Conversions &
    (b) Gas Network Shutdowns \& Grid Upgrades\\
    \end{tabular}
    \vspace{-0.43cm}
    \caption{The performance of \emph{network-aware} strategy at varying fractions of reference gas network maintenance cost.}
    \label{fig:results2}
    \vspace{0.2cm}
\end{figure*}

%

\vspace{-0.15cm}
\subsection{Network-aware Transition}\label{sec:evalNaive} 

We first analyze the performance of the \emph{network-aware} and \emph{network-oblivious} transition strategies for a given budget. 
Figure~ \ref{fig:naiveVSnetwork} shows that our strategy outperforms the baseline across all metrics. 
At 10\% of the benchmark budget, our \emph{network-aware} strategy not only achieves 55.1\% higher carbon reduction than baseline (11.23\% vs 7.24\%), it also exceeds the expected emission reductions for the budget by 12.3\%. 
While both strategies convert almost the same number of households (3.61\% vs. 3.57\%), the \emph{network-aware} strategy shuts down roughly 10 times more pipelines (6.00\% vs 0.62\%). 
This happens because our strategy selects entire blocks and neighborhoods at once leading to higher gas network shutdown, which eliminates maintenance cost for the utility in perpetuity -- these savings can be reinvested into home transitions, with the effect of achieving higher \emph{overall} carbon reduction. 
Finally, our \emph{network-aware} strategy also requires upgrading roughly 64\% less transformers than baseline (3.48\% vs. 9.22\%) post-transition. 

Comparing the \emph{network-aware} and \emph{network-oblivious} transition strategies for 20\% of the benchmark budget, our \emph{network-aware} strategy achieves 89.3\% higher carbon reduction than the baseline (21.72\% vs 11.47\%), and also exceeds the expected emission reductions for the budget by 8.6\%.  The \emph{network-aware} strategy converts 67.1\% more households (7.83\% vs. 4.68\%), shuts down roughly 17 times more pipelines (10.85\% vs 0.62\%), and requires upgrading roughly 74\% less transformers (6.23\% vs. 24.36\%) post-transition. 

This experiment yields another interesting observation. 
Our \emph{network-aware} strategy shuts down a larger segment of the gas network (\emph{higher savings}) and requires upgrading lesser number of transformers (\emph{lower cost}) than the \emph{network-oblivious} strategy, but converts the same number of homes. 
Where does the extra money go and why are carbon emission reductions high?
A deep dive into the results reveals that the \emph{network-aware} strategy generally picks bigger homes that are situated in less dense areas of the city. 
This allows shutting down a larger portion of the network for the same number of households transitioned. 
However, the extra savings are spent on installing larger heat pumps for the bigger homes since their usage is higher than the median usage. 
This also leads to higher carbon reduction for the same number of homes.

This observation is validated by a visual inspection of Figure~\ref{fig:visualResults}(b) and Figure~\ref{fig:visualResults}(c). 
The \emph{network-oblivious} strategy converts households that are scattered uniformly throughout the city, while \emph{network-aware} transitions households in big blocks and neighborhoods. 

\textbf{\emph{Key takeaway.}} \emph{A network-aware strategy yields higher savings by shutting down more of the gas network, saves cost on transformer upgrades, and achieves higher carbon emission reductions by transitioning bigger, and marginally higher number of, households.}

\vspace{0.15cm}
\noindent\textbf{Impact of budget.}
The extent of carbon reductions and gas network shutdown for a transition strategy depends on its budget.
While our strategy outperforms the expected carbon reductions at 10\% of the benchmark budget, we want to examine if the advantage holds in the case of a larger or smaller budget than our initial analysis. 
Figure \ref{fig:results1}(a) shows that \emph{network-aware} strategy outperforms the budget-estimated reductions in each tested scenario, by a minimum of 8\% (at 5\% \& 15\% budgets) and a maximum of 11.2\% (at 10\% budget). 
We also achieve a super-linear increase in households transitioned as budget increases -- at least 1.8\% increase in transitioned households for every 5\% budget increase.

Figure \ref{fig:results1}(b) demonstrates that a \emph{network-aware} strategy is able to restrain the increase in transformer upgrades required as budget increases. For example, we observe a mere 0.9\% increase in upgraded transformers as budget increases from 15\% to 20\%. 
However, the extent of gas network shutdown does not scale well with budget -- a 10\% budget yields 6\% gas network shutdown while a 20\% budget results in less than 12\%, 10.8\% to be exact, gas network shutdown.  As the budget increases, our framework selects new neighborhoods which are comparatively shorter and denser -- if the budget increased further and worked through the dense downtown area, which has short neighborhood gas lengths, we would see larger sections of the network being decommissioned.

\textbf{\emph{Key takeaway.}} \emph{A network-aware strategy scales well with the increase in budget and consistently outperforms the expected carbon emission reductions and the number of households transitioned.}



\vspace{0.15cm}
\noindent\textbf{Impact of gas network maintenance cost.}
The cost of gas network maintenance is another important parameter in our analysis. 
We use the utility's yearly expenditure on gas network maintenance as our reference cost (see~\S\ref{sec:cost-models} for details), and scale it in the range of $0.5-2.0$ for our analysis. 
The reduction in maintenance cost can serve as a proxy for the time horizon for savings -- $0.5$ corresponds to $6$ months and $2$ corresponds to $2$ years. 
The maintenance cost can also serve as a coarse method for simulating different ``states of repair'' for the gas network -- an old gas network would incur higher maintenance expenditure than a new one~\cite{Gallagher:2015}. 





Figure~\ref{fig:results2}(a) demonstrates that a higher maintenance cost enables the \emph{network-aware} strategy to achieve higher carbon reductions, but they do not scale well. 
We observe a small (0.1\%) increase in carbon reductions as maintenance cost increases by 50\% from 1.5x to 2.0x. 
The increase in the number of households converted also tails off with increased maintenance cost -- 2.8\% increase from 1.0x to 1.5x versus 1.1\% increase from 1.5x to 2.0x.  This happens because, under a fixed budget, the extra savings from gas line shutdowns are spent on upgrading transformers, as shown in Figure~\ref{fig:results2}(b). 

A deep dive into the results reveals that the set of households transitioned at higher maintenance costs is bigger but is not always a super-set of the households transitioned at a lower cost. 
This results in diminished returns on carbon reductions as a more uniform set of households, including less carbon-intensive homes, get transitioned due to the increased cost of maintenance.

\textbf{\emph{Key takeaway.}} \emph{A network-aware strategy transitions more households, shuts-down a larger gas network, and upgrades more transformers as the maintenance cost increases. However, gas maintenance cost does not have a significant impact on carbon emission reductions.}
\vspace{-1.5em}

\vspace{-0.25cm}
\subsection{Equity-Aware Transition}
Our \emph{equity-aware} optimization framework, described in~\S\ref{sec:equityDef}, provides a mechanism to allocate a specific budget for each socioeconomic group under consideration. 
We conduct our analysis for a scenario where each income tract gets one third of the budget. 
However, we acknowledge that the definition of \emph{equity} can significantly vary depending upon the context and reiterate that our framework supports any split across groups. 
We set the total budget to 10\% of the benchmark and use reference gas maintenance cost. 

\begin{figure}[t]
    \vspace{0.4cm}
    \includegraphics[width=0.9\linewidth]{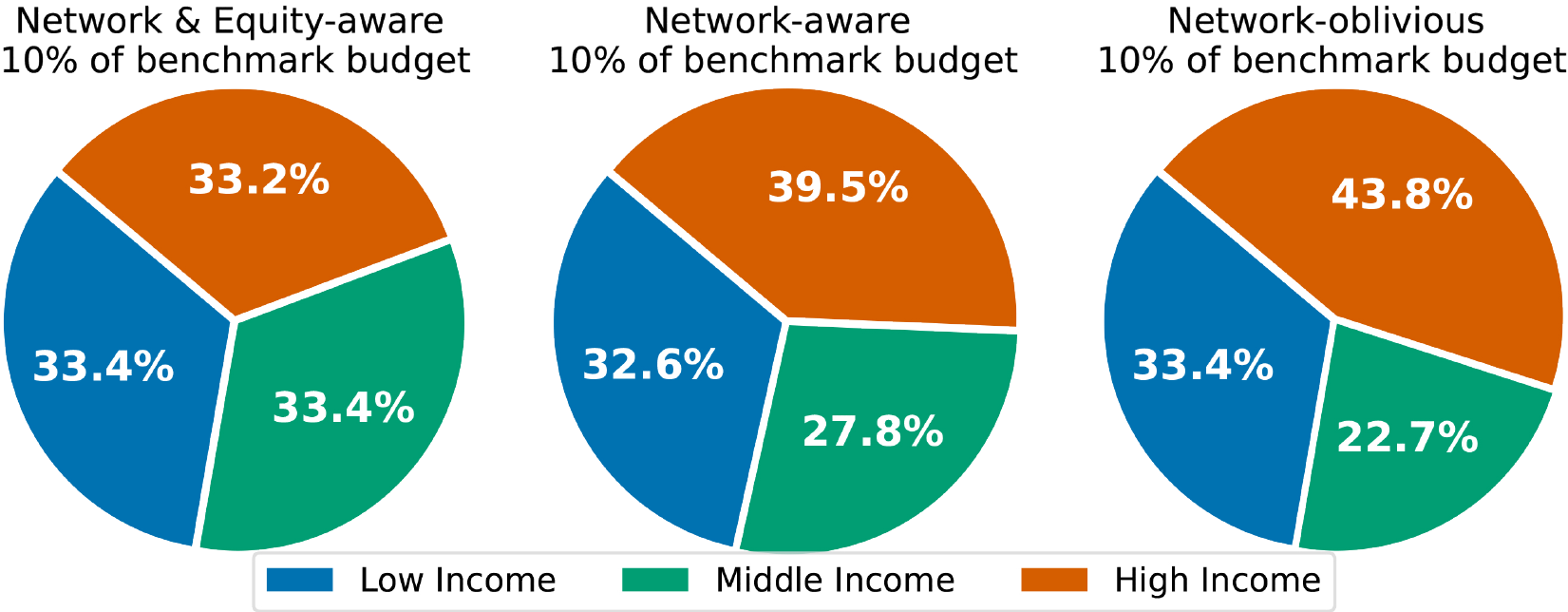}
    
    \caption{Our \emph{equity-aware} strategy spends exactly one-third of the budget on each income tract (left) while the \emph{network-aware} (middle) and \emph{network-oblivious} (right) strategies favour larger, high-income households}
    \label{fig:equityDist}
    \vspace{-0.5cm}
\end{figure}

Figure \ref{fig:equityDist} shows the distribution of budget spent on transitioned homes across different income tracts. 
The \emph{equity-aware} strategy achieves the desired distribution, while both \emph{network-aware} and \emph{network-oblivious} strategies fail, especially for the middle income tract. 
The \emph{network-oblivious} strategy simply picks the highest emitters generally belonging to high income tract. 
Our \emph{network-aware} strategy picks sparsely located homes that require few transformer upgrades and save the most from gas maintenance costs.
These households are generally located near the edge of the network and belong to middle- and high-income tracts.



A visual inspection of Figure~\ref{fig:visualResults}(c) and Figure~\ref{fig:visualResults}(d) validates our observations.
The \emph{equity-aware} framework leaves out some of the high-income households (red shaded areas) and picks more low- and middle-income households (green shaded areas) to achieve an equitable spread across income tracts. 

\begin{figure}[t]
    \includegraphics[width=\linewidth]{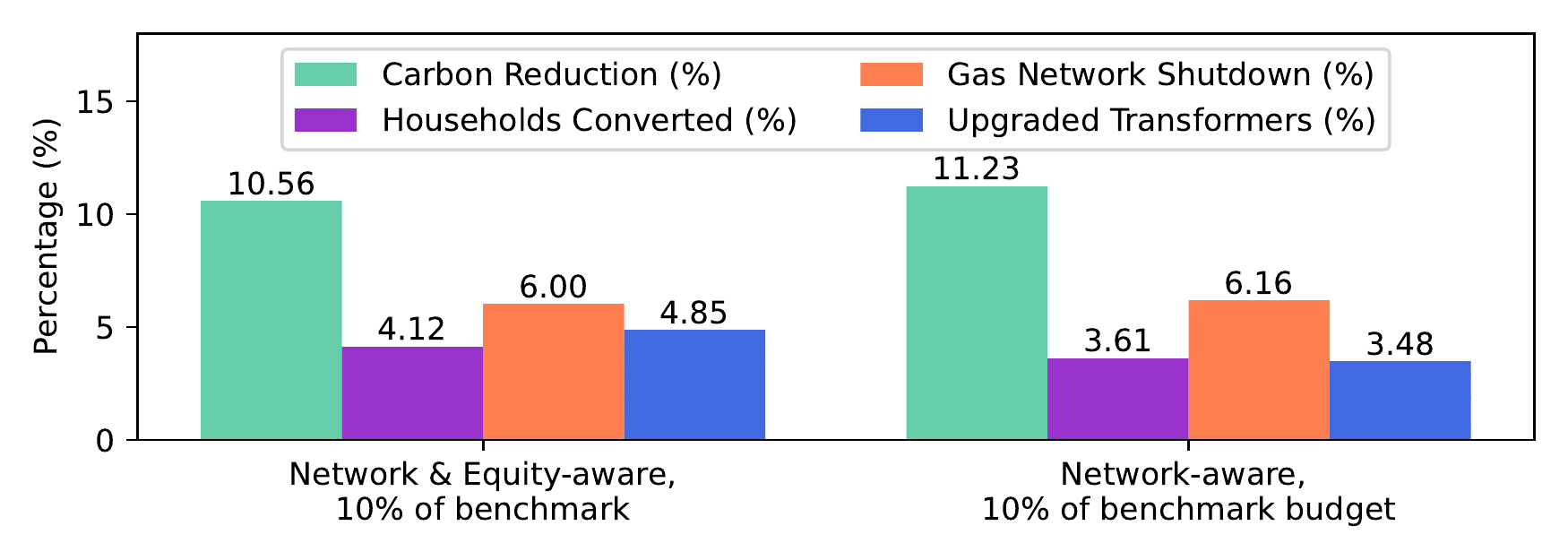}
    \vspace{-0.6cm}
    \caption{Our \emph{equity-aware} strategy transitions more households at the expense of a marginal decrease in carbon reductions, but outperforms budget-based reduction estimates.}
    \label{fig:equityComp}
    \vspace{-0.6cm}
\end{figure}

Our \emph{equity-aware} approach results in expected trade-offs across other metrics, as shown in Figure~\ref{fig:equityComp}.
The \emph{equity-aware} strategy achieves 5.9\% less carbon reductions as it trades high-income larger homes for slightly smaller middle-income homes, a desired outcome from an equity perspective.  
Since bigger homes also cost more to transition, \emph{equity-aware} approach is able to convert 14.3\% more households than the \emph{network-aware} strategy.
The \emph{equity-aware} strategy upgrades almost 40\% more transformers to accommodate the larger number of transitioned households, while the gas network shutdown remains the same across the strategies.

\textbf{\emph{Key takeaway.}} \emph{An equity-aware strategy achieves slightly less carbon reductions, but transitions more households to achieve an equitable spread of budget across different income tracts.}

\section{Related Work}
\label{sec:related}

In this section, we examine prior work on leveraging heat pumps for residential heating decarbonization, network-aware heating decarbonization, and equity-aware energy transition efforts.

\noindent{\bf Heat pumps for heating decarbonization.}
There has been numerous studies that investigate the use of Air Source Heat Pumps (ASHP) as an alternative to gas-based heating, especially in cold climates~\cite{ashp-review-2020, ashp-potential-2016, ashp-feasibility-china-2020, ashp-analysis-china-2019}. 
\citet{ashp-review-2020} present a systemic review of ASHP field studies.
\citet{ashp-potential-2016} evaluate the potential of ASHP as an alternative to fossil fuels and analyze their socio-economic impact for Ireland.
\citet{ashp-feasibility-china-2020} analyze the feasibility and performance of hybrid ASHP in severe cold region of China.
Another body of work specifically explores the decarbonization potential of heat pumps~\cite{kaufman2019decarbonizing,  zhang2019decarbonizing, kontu2019introducing, neirotti2020towards}. 
\citet{kaufman2019decarbonizing} analyzes the cost-competitiveness and decarbonization potential of ASHP in US.
Similar studies have been performed for Beijing~\cite{zhang2019decarbonizing}, entire China~\cite{ashp-analysis-china-2019}, Finland~\cite{kontu2019introducing}, and other countries across Europe~\cite{neirotti2020towards}. 

While these studies help design high-level policies for transitioning to heat pumps and decarbonization, they do not provide granular information that can guide how such a transition should be carried out for a given gas and electricity network. 
Our work follows such feasibility studies to inform how transitioning to heat pumps should be carried out at a city-scale while respecting the constraints of both the gas network and the electric grid.

\noindent{\bf Network-aware heating decarbonization.}
Prior work also investigates the impact of transitioning to electric heat pumps on the electric grid~\cite{waite2020electricity, optimal-g-e-planning-2022, gis-based-planning-2020, gas-elec-depend-2020}.
\citet{gis-based-planning-2020} evaluate different heat pump scenarios including electric, gas, and hybrid heat pumps. However, their analysis is limited to 100\% transition scenario, contains only 12 buildings, constructs gas and electric load profiles synthetically, and only considers the effect on power lines ignoring key components such as transformers. 
\citet{optimal-g-e-planning-2022} did a similar analysis using a small 24-node synthetic electric and gas network, and load profiles. 
\citet{gas-elec-depend-2020} explore the inter-dependencies of gas and electric networks for a small segment of a German city. 
They conduct a multi-agent simulation to model the retrofit decisions of building owners in response to changing gas and electricity charges as gas consumers gradually defect. 
They do not look at the problem from the utility perspective that wants to identify the optimal part of gas network for transition and also ignore equity.

Our work differs from prior work in three key ways. 
First, we take a utility perspective that wants to identify the optimal segments of network for transition and identify relevant upgrades in power grid under budgetary constraints. 
Second, we devise an equity-aware approach that ensures that the benefits of transition are distributed across all income groups. 
Third, we use electricity and gas demand profiles from actual homes situated in the city under-consideration.

\noindent{\bf Equity-aware heating decarbonization.}
There have been multiple studies that investigate inequity in the energy transition and show that lower-income and marginalized communities are negatively impacted in the process~\cite{banzhaf2019environmental, ross2018high}. 
Multiple studies suggest incorporating equity into the energy transition policies~\cite{drehobl2016lifting, drehobl2020high}.
\citet{WamburuGrazierIrwinCragoShenoy:2022} propose a decarbonization strategy that considers equity while selecting buildings for heat pump retrofits. 

As discussed, these studies either highlight the existing inequity, propose incorporating equity, or include equity in coarse retrofit decisions. 
None of these studies take a network-aware, budget-constrained utility's perspective while incorporating equity.

\section{Conclusion}
\label{sec:conclusion}

In this paper, we presented a network- and equity-aware optimization framework to optimize carbon reduction under budgetary constraints while transitioning from gas-based heating to electric heat pumps from a utility's perspective. 
We evaluated our framework using real natural gas and electricity usage data from a city in New England.  
Leveraging knowledge of the gas network's topology, our framework achieves 55\% more carbon reduction than prior \emph{network-oblivious} methods by targeting specific segments of the gas network. 
Furthermore, our equity analysis shows that the \emph{network-oblivious} and the \emph{network-aware} strategies fail to achieve equitable distribution, while our \emph{equity-aware} approach achieves an equitable outcome which preserves the carbon reduction benefits of the \emph{network-aware} strategy.

\begin{acks}
We thank the anonymous reviewers for their insightful comments. This research is supported by NSF grants 1908298, 2020888, 2021693, 2045641, 2105494, and 2136199.
\end{acks}

\balance

\bibliographystyle{ACM-Reference-Format}
\bibliography{main}

\end{document}